\pdfoutput=1
\documentclass[aps,pre,onecolumn,superscriptaddress,notitlepage]{revtex4-1}
\usepackage{subfigure}
\usepackage{amsmath,graphicx,amssymb}
\usepackage{multirow,color,float}
\usepackage{url}
\usepackage{verbatim}
\usepackage{bm}

\newcommand{\Ceps}{C_\varepsilon}
\newcommand{\Cinf}{C_{\varepsilon,\infty}}
\newcommand{\Cepsu}{C_{\varepsilon,u}}

\newcommand{\Rl}{R_{\lambda}}
\newcommand{\vep}{\varepsilon}

\newcommand{\sigmapm}{\sigma_\pm}
\newcommand{\etal}{\textit{et al.}}
\renewcommand{\vec}[1]{\bm{#1}}
\newcommand{\fvec}[1]{\hat{\bm{#1}}}
\newcommand{\vx}{\vec{x}}

\renewcommand{\vr}{\vec{r}}

\newcommand{\vu}{\vec{u}}

\newcommand{\vb}{\vec{b}}

\newcommand{\vk}{\vec{k}}

\newcommand{\zpm}{\vec{z}^\pm}
\newcommand{\zmp}{\vec{z}^\mp}

\newcommand{\blue}[1]{{#1}}
\newcommand{\beq}{\begin{equation}}
\newcommand{\eeq}{\end{equation}}
\begin{document}
\title{Reynolds-number dependence of the dimensionless dissipation
       rate in homogeneous magnetohydrodynamic turbulence}

\author{Moritz Linkmann}
\email[]{linkmann@roma2.infn.it}
\affiliation{Department of Physics \& INFN, University of Rome Tor Vergata, Via della Ricerca Scientifica 1, 00133 Rome, Italy}
\affiliation{SUPA, School of Physics and Astronomy, University of Edinburgh, Peter Guthrie Tait Road, EH9 3FD, UK}

\author{Arjun Berera}
\affiliation{SUPA, School of Physics and Astronomy, University of Edinburgh, Peter Guthrie Tait Road, EH9 3FD, UK}

\author{Erin E.~Goldstraw}
\affiliation{School of Mathematics and Statistics, University of St.~Andrews, KY16 9SS, UK}
\affiliation{SUPA, School of Physics and Astronomy, University of Edinburgh, Peter Guthrie Tait Road, EH9 3FD, UK}

\date{\today}

\begin{abstract} 
This paper examines the behavior of the dimensionless dissipation rate $\Ceps$
for stationary and nonstationary magnetohydrodynamic (MHD) turbulence in
presence of external forces. By combining with previous studies for freely
decaying MHD turbulence, we obtain here both the most general model equation
for $\Ceps$ applicable to homogeneous MHD turbulence and a comprehensive
numerical study of the Reynolds number dependence of the dimensionless total
energy dissipation rate \blue{at unity magnetic Prandtl number}.  
We carry out a series of medium to high resolution
direct numerical simulations of mechanically forced stationary MHD turbulence
in order to verify the predictions of the model equation for the stationary
case. Furthermore, questions of nonuniversality are discussed in terms of the
effect of external forces as well as the level of cross- and magnetic helicity.
The measured values of the asymptote $\Cinf$ lie between $0.193 \leqslant \Cinf
\leqslant 0.268$ for free decay, where the value depends on the initial level
of cross- and magnetic helicities. In the stationary case we measure $\Cinf =
0.223$.
\end{abstract}

\maketitle

\section{Introduction}
The dynamics of conducting fluids is relevant to many areas in geo- and astrophysics as well as
in engineering and industrial applications. Often the flow is turbulent, and the 
interaction of the turbulent flow with the magnetic field leads to considerable complexity. 
Being a multi-parameter problem, techniques that have been successfully applied to 
turbulence in nonconducting fluids sometimes fail to deliver unambiguous predictions
in magnetohydrodynamic (MHD) turbulence. This  
concerns e.g.~the prediction of inertial range scaling exponents by extension of 
Kolmogorov's arguments \cite{Kolmogorov41a} to MHD, and considerable effort 
has been put into the further understanding of inertial range cascade(s) in MHD 
turbulence \cite{Iroshnikov64,Kraichnan65a,Goldreich95,Boldyrev05a,Boldyrev06,
Beresnyak06,Mason06,Gogoberidze07}.           
The difficulties are partly due to the many different configurations that can arise
in MHD turbulence because of e.g.~anisotropy, 
different levels of vector field correlations, different 
values of the dissipation coefficients and
different types of external forces, and as such are  
connected to the question of universality in MHD turbulence 
\cite{Dallas13a,Dallas13b,Wan12,Schekochihin08,Mininni11,
Grappin83,Pouquet08,Beresnyak11,Boldyrev11,Grappin10,Lee10,Servidio08}.
The behavior of the (dimensionless) dissipation
rate is representative of this problem, in the sense that the aforementioned properties of MHD turbulence 
influence the energy transfer across the scales, {\em i.e.}~the cascade dynamics 
\cite{Frisch75,Pouquet76,Pouquet78,Biskamp93,Dallas13b,Alexakis13}, and thus the amount of
energy that is eventually dissipated at the small scales.

The behavior of the total dissipation rate in a turbulent non-conducting fluid is 
a well-studied problem. As such it has been known for a long time that
the total dissipation rate in both stationary and freely decaying homogeneous
isotropic turbulence tends to a constant value with increasing Reynolds number following
a well-known characteristic curve 
\cite{Sreenivasan84,Sreenivasan98,McComb14a,McComb15a,Yeung15,Donzis16}.
\blue{ For statistically steady isotropic turbulence this curve can be
approximated by the real-space stationary energy balance equation, where the
asymptote is connected to the maximal inertial flux of kinetic energy
\cite{McComb15a}.  }
The corresponding problem in MHD has received much less attention, however,
recent numerical results for freely decaying MHD turbulence \blue{at unity
magnetic Prandtl number} report similar behavior.  Mininni and Pouquet
\cite{Mininni09} carried out direct numerical simulations (DNSs) of freely
decaying homogeneous MHD turbulence without a mean magnetic field, showing that
the temporal maximum of the total dissipation rate $\vep(t)$ became independent
of Reynolds number at a Taylor-scale Reynolds number $\Rl$ (measured at the
peak of $\vep(t)$) of about 200.  Dallas and Alexakis \cite{Dallas14b} measured
the dimensionless dissipation rate $\Ceps$ also from DNS data for free decay
for random initial fields with strong correlations between the velocity field
and the current density. Again, it was found that $\Ceps \to const.$ with
increasing Reynolds number.  Interestingly, a comparison with the data of
Ref.~\cite{Mininni09} showed that the approach to the asymptote was slower than
for the data of Ref.~\cite{Mininni09}, suggesting an influence of the level of
certain vector field correlations on the approach to the asymptote.  A
theoretical model for dissipation rate scaling in freely decaying MHD
turbulence was put forward recently \cite{Linkmann15a} based on the von
K\'arm\'an-Howarth energy balance equations (vKHE) in terms of Els\"{a}sser
fields \cite{Politano98}. \blue{For unity magnetic Prandtl number it} predicts
the dependence of $\Ceps$ on a generalized Reynolds number \blue{$R_- \equiv
z^-L^+/(\nu + \mu)$, with $z^-$ denoting the root-mean-square value of one
Els\"asser field, $L^+$ the integral scale corresponding to the other
Els\"asser field, while $\nu$ and $\mu$ are the kinematic viscosity and the
magnetic resistivity, respectively.   The model equation has the following
form}
\beq
\label{eq:approx}
\Ceps = \Cinf + \frac{C}{R_-} + \frac{D}{R_-^2} + O(R_-^{-3}) \ ,
\eeq
where $C$ and $D$ are time-dependent coefficients depending on several parameters, 
which themselves depend on the magnetic, cross- and kinetic helicities.
The predictions
of this equation were subsequently tested against data obtained from medium to high
resolution DNSs of freely decaying homogeneous MHD turbulence leading to a very good agreement between
theory and data. 

In summary, there is compelling numerical and theoretical evidence for finite
dissipation in freely decaying MHD turbulence \blue{at least for unity magnetic
Prandtl number $Pm = \nu/\mu$}, while so far no systematic results for the stationary case
have been reported.  In this paper we extend the derivation carried out in
Ref.~\cite{Linkmann15a} to include the effects of external forces and we
present the first systematic study of dissipation rate scaling for stationary
MHD turbulence. \blue{ In order to be able to test the model equation against
DNS data  for a large range of generalized Reynolds numbers, we concentrate on
the case $Pm = 1$.} The most general form of
Eq.~\eqref{eq:approx} for nonstationary flows with large-scale external forcing
is derived, which can be applied to freely decaying and stationary flows by
setting the corresponding terms to zero. This generalization of
Eq.~\eqref{eq:approx}  is the first main result of the paper, it is applicable
to both freely decaying and stationary MHD turbulence. It implies that the
dissipation rate of total energy is finite in the limit $R_- \to \infty$ in
analogy to hydrodynamics, and highlights the dependence of the coefficients $C$
and $D$ on the external forces. As such, Eq.~\eqref{eq:approx} predicts
nonuniversal values of the asymptotic value $\Cinf$ of the dimensionless
dissipation rate in the infinite Reynolds number limit and of the approach to
the asymptote for a variety of MHD flows.  The resulting theoretical
predictions for the stationary case are compared to DNS data for stationary MHD
turbulence for three different types of mechanical forcing while the results
for the freely decaying case \cite{Linkmann15a} are reviewed for completeness.
The DNS data shows good agreement with Eq.~\eqref{eq:approx} and the different
forcing schemes have no measurable effect on the values of the coefficients in
Eq.~\eqref{eq:approx}. The measured values of $\Cinf$ lie between $0.193
\leqslant \Cinf \leqslant 0.268$ for free decay, where the value depends on the
initial level of cross- and magnetic helicities. In the stationary case we
measure $\Cinf = 0.223$.  
 
This paper is organized as follows. We begin by reviewing the formulation of
the MHD equations in terms of Els\"asser fields in Sec.~\ref{sec:definitions}
where we introduce the basic quantities we aim to study in both formulations of
the MHD equations.  In Section \ref{sec:derivation} we extend the derivation
put forward in Ref.~\cite{Linkmann15a} to nonstationary MHD turbulence.  The
model equation is verified against DNS data for statistically steady MHD
turbulence and the comparison to data for freely decaying MHD turbulence
presented in Ref.~\cite{Linkmann15a} is reviewed in Sec.~\ref{sec:numerics},
where special emphasis is given to the question of nonuniversality of MHD
turbulence in the context of external forces and the level of cross- and
magnetic helicities.  Our results are summarized and discussed in the context
of related work in hydrodynamic and MHD turbulence in
Sec.~\ref{sec:conclusions}, where we also outline suggestions for further work.
\\

\section{The total dissipation in terms of Els\"asser fields}
\label{sec:definitions}
In this paper we consider statistically homogeneous MHD turbulence in the absence 
of a background magnetic field. The flow is taken to be incompressible, leading to the following
set of coupled partial differential equations
\begin{align}
\label{eq:momentum}
\partial_t \vec{u}&= - \frac{1}{\rho}\nabla P -(\vec{u}\cdot \nabla)\vec{u}
 + \frac{1}{\rho}(\nabla \times \vec{b}) \times \vec{b} + \nu \Delta \vec{u} +  \vec{f}_u  \ , \\
\label{eq:induction}
\partial_t \vec{b}&= (\vec{b}\cdot \nabla)\vec{u}-(\vec{u}\cdot \nabla)\vec{b} + \mu \Delta \vec{b} + \vec{f}_b\ , \\
\label{eq:incompr}
&\nabla \cdot \vec{u} = 0 \ \ \mbox{and} \ \  \nabla \cdot \vec{b} = 0 \ ,  
\end{align}
where $\vec{u}$ denotes the velocity field, $\vec{b}$ the magnetic
induction expressed in Alfv\'{e}n units, 
$\nu$ the kinematic viscosity, $\mu$ the
magnetic resistivity, $P$ the thermodynamic pressure, $\vec{f}_u$ and $\vec{f}_b$ are
external mechanical and electromagnetic forces, which may be present,
and $\rho$ denotes the density which is set to unity for convenience.
Equations \eqref{eq:momentum}-\eqref{eq:incompr} are considered on a three-dimensional
domain $\Omega$, which due to homogeneity can either be the full space $\mathbb{R}^3$ 
or a subdomain $[0,L_{box})^3$ with periodic boundary conditions.
The MHD equations \eqref{eq:momentum}-\eqref{eq:incompr} 
can be formulated more symmetrically using Els\"asser variables
$\vec{z^\pm}=\vec{u} \pm \vec{b}$ \cite{Elsasser50}
\begin{align}
\partial_t \zpm&= - \frac{1}{\rho}\nabla \tilde{P} -(\zmp\cdot \nabla)\zpm 
+ (\nu + \mu) \Delta \zpm + (\nu - \mu) \Delta \zmp + \vec{f}^\pm  \ , \\
&\nabla \cdot \zpm = 0 \ ,  
\label{eq:elsasser}
\end{align}
where $\vec{f}^\pm = \vec{f}_u \pm \vec{f}_b$ and
the pressure $\tilde{P}$ consists of the sum of the thermodynamic pressure $P$ and the
magnetic pressure $\rho|\vb|^2/2$. Which formulation of the MHD equations is chosen
often depends on the physical problem, for some problems the Els\"asser formalism
is technically convenient, while the formulation using the primary fields $\vu$ and $\vb$
facilitates physical understanding.
The ideal invariants total energy $E(t)$, cross-helicity $H_c(t)$ 
and magnetic helicity $H_m(t)$ are given 
in the respective formulations of the MHD equation by
\begin{align}
E(t) &= 
\frac{1}{2} \int_\Omega d \vk \ \langle |\fvec{u}(\vk,t)|^2 + |\fvec{b}(\vk,t)|^2 \rangle 
        = \frac{1}{4}\int_\Omega d \vk \ \langle |\fvec{z}^+(\vk,t)|^2+|\fvec{z}^-(\vk,t)|^2 \rangle \ , \\
H_{c}(t)& = \int_\Omega d \vk \ \langle \fvec{u}(\vk,t) \cdot \fvec{b}(-\vk,t) \rangle 
        = \frac{1}{4}\int_\Omega d \vk \ \langle |\fvec{z}^+(\vk,t)|^2-|\fvec{z}^-(\vk,t)|^2 \rangle \ , \\
H_{m}(t)& = \int_\Omega d \vk \ \langle \fvec{a}(\vk,t) \cdot \fvec{b}(-\vk,t) \rangle 
        = \frac{1}{4}\int_\Omega d \vk \ 
\left \langle \left[\frac{i\vk}{k^2} \times (\fvec{z}^+(\vk,t)-\fvec{z}^-(\vk,t))\right] \cdot (\fvec{z}^+(-\vk,t)-\fvec{z}^-(-\vk,t))\right \rangle \ ,
\label{eq:els_hmag}
\end{align}
with $\fvec{b}$, $\fvec{u}$ and $\fvec{z}^\pm$ denoting the respective Fourier transforms of the magnetic, 
velocity and Els\"asser fields, while $\fvec{a}$ is the Fourier transform of the magnetic vector potential $\vec{a}$. 
The angled brackets indicate an ensemble average.
\blue{Equation \eqref{eq:els_hmag} is gauge-independent as shown in Appendix \ref{app:els_hmag}.}
\\

We now motivate the use of the Els\"asser formulation for the study of the dimensionless dissipation coefficient in MHD. 
In hydrodynamics, the dimensionless dissipation coefficient
$\Cepsu$ is defined in terms of the Taylor surrogate expression
for the total dissipation rate, 
$U^3/L_u$, 
where $U$ denotes the root-mean-square (rms) value of the velocity field and
$L_u$ the integral scale defined with respect to the 
velocity field, as
\beq
\Cepsu \equiv \vep_{kin} \frac{L_u}{U^3} \ .
\eeq  
However, in MHD there are several quantities that
may be used to define an MHD analogue to the Taylor 
surrogate expression, such as the rms value $B$ of the 
magnetic field, one of the different length scales 
defined with respect to either $\vec{b}$ or $\vec{u}$, or the
total energy. 

Since the total dissipation in MHD turbulence should be related to the 
flux of total energy through different scales, one may think of defining 
a dimensionless dissipation coefficient for MHD in terms of the total 
energy. However, this would lead to a nondimensionalization of 
the hydrodynamic transfer term 
$\vec{u}\cdot(\vec{u}\cdot \nabla) \vec{u}$ with a magnetic quantity.
This can be seen by considering the analog of the von K\'arm\'an-Howarth energy balance equation in real space
\cite{Chandrasekhar51} stated here for the case of free decay
\begin{align}
-d_t E(t)=\vep(t) = &  -\partial_t (B_{LL}^{uu}(r,t)+B_{LL}^{bb}(r,t))
 +\frac{3}{2r^4} \partial_r \left(\frac{r^4}{6} B_{LLL}^{uuu}(r,t) +r^4 C_{LLL}^{bbu}(r,t) \right)  \nonumber \\
&  \ \ \ + \frac{6}{r} C^{bub}(r,t) + \frac{1}{r^4} \partial_r \left (r^4 \partial_r (\nu B_{LL}^{uu}(r,t)+ \mu B_{LL}^{bb}(r,t)) \right) \ , 
\label{eq:ub_energy}
\end{align}
where $B_{LL}^{uu}$, $B_{LL}^{bb}$ and $B_{LLL}^{uuu}$ are the longitudinal structure functions, 
$C_{LLL}^{bbu}$ the longitudinal correlation function and $C^{bub}$ another correlation function. The 
longitudinal structure and correlation functions are given by
\begin{align}
B_{LL}^{uu}(r,t)&= \langle (\delta u_L(\vec{r},t))^2  \rangle \ , \\ 
B_{LL}^{bb}(r,t)&= \langle (\delta b_L(\vec{r},t))^2 \rangle  \ , \\
B_{LLL}^{uuu}(r,t)&= \langle (\delta u_L(\vec{r},t))^3 \rangle  \ , \\ 
C_{LLL}^{bbu}(r,t)&= \langle u_L(\vec{x},t) b_L(\vec{x},t) b_L(\vec{x} + \vec{r},t) \rangle \ ,  
\end{align}
where $r=|\vr|$ and
$v_L = \vec{v}\cdot\vec{r}/r$ denotes the longitudinal component of a vector field $\vec{v}$,
that is its component parallel to the displacement vector $\vec{r}$, and
\beq
\delta v_L(\vr) = [\vec{v}(\vec{x}+\vec{r})-\vec{v}(\vec{x})]\cdot \frac{\vec{r}}{r} \ ,
\eeq
its longitudinal increment. The function $C^{bub}$ is defined through the third-order 
correlation tensor
\begin{align}
C_{ij,k}^{bub}(\vec{r},t)&= 
\langle (u_i(\vec{x}) b_j(\vec{x})-b_i(\vec{x}) u_j(\vec{x})) b_k(\vec{x} + \vec{r}) \rangle = 
C^{bub}(r,t)\left(\frac{r_j}{r} \delta_{ik} - \frac{r_i}{r} \delta_{jk}  \right)  . \  
\end{align}
As can be seen from their respective definitions, the functions $C_{LLL}^{bbu}$
and $C^{bub}$ scale with $B^2U$ while the function $B_{LLL}^{uuu}$ scales with
$U^3$. If Eq.~\eqref{eq:ub_energy} were to be nondimensionalized with 
respect to the total energy 
then the purely hydrodynamic term $B_{LLL}^{uuu}$ would be scaled partially 
by a magnetic quantity. \\

This problem of inconsistent nondimensionalization can be avoided by working with Els\"asser 
fields, which requires an expression for the total dissipation rate $\vep(t)$ 
in terms of Els\"asser fields. 
The total rate of energy dissipation in MHD turbulence is given by
the sum of Ohmic and viscous dissipation 
\beq
\varepsilon(t) = \varepsilon_{mag}(t) + \varepsilon_{kin}(t) \ ,
\eeq
where 
\begin{align}
\vep_{mag}(t) & = \mu \int_\Omega d\vec{k} \ k^2 \langle |\fvec{b}(\vk,t) |^2 \rangle \ , \\
\vep_{kin}(t) &=  \nu \int_\Omega d\vec{k} \ k^2 \langle |\fvec{u}(\vk,t) |^2 \rangle \ .
\end{align}
Similarly, the total dissipation rate can be decomposed into its respective 
contributions from the Els\"asser dissipation rates
\beq
\varepsilon(t) = \frac{1}{2} \big (\varepsilon_{+}(t) + \varepsilon_{-}(t) \big ) \ ,
\eeq
where the Els\"{a}sser dissipation rates are defined as 
\beq 
\vep^\pm(t) = \nu_+ \int_\Omega d\vec{k}  \ k^2 \langle |\fvec{z}^\pm(\vk,t) |^2 \rangle 
            + \nu_- \int_\Omega d\vec{k}  \ k^2 \langle \fvec{z}^\pm(\vk,t) \cdot \fvec{z}^\mp(-\vk,t) \rangle 
\ , 
\eeq
with $\nu_\pm = (\nu \pm \mu)$. 
The total dissipation rate relates to the sum of the Els\"{a}sser
dissipation rates 
\beq
\vep^+(t) + \vep^-(t) = \vep(t) +\vep_{H_{c}}(t) + \vep(t) -\vep_{H_{c}}(t) = 2 \vep(t) \ ,
\eeq
where 
the cross-helicity dissipation rate $\vep_{H_{c}}$ is given by 
\beq
\vep_{H_{c}}(t) = \frac{1}{2} \big(\vep^+(t) - \vep^-(t)\big) \ .
\eeq
Since this paper is concerned with \blue{both stationary and} nonstationary 
flows, the total energy input rate 
$\iota$ must also be considered. Similar to the dissipation rate, the input rate can be split up 
into either kinetic and magnetic contributions or the Els\"asser contributions $\iota^\pm(t)$ 
\begin{align}
\label{eq:diss_ub}
\iota(t) &= \iota_{mag}(t) + \iota_{kin}(t) \\
\label{eq:diss_zz}
\iota(t)&=\frac{1}{2}\big(\iota^+(t) + \iota^-(t) \big) \ .
\end{align}
The latter equation can be rewritten as
\beq
\iota^+(t) = \iota(t) + \frac{1}{2}\left( \iota^+(t) - \iota^-(t) \right) = \iota(t) + \iota_{H_c}(t) \ ,
\eeq
where $\iota_{H_c}$ denotes the input rate of the cross-helicity.

\section{Derivation of the equation} 
\label{sec:derivation}
Since the total dissipation rate can be expressed either in terms of the Els\"{a}sser 
fields or the primary fields $\vec{u}$ and $\vec{b}$, it should be possible to describe it also 
by the vKHE for $\vec{z}^\pm$ \cite{Politano98}. For the freely decaying case 
no further complication arises as the rate of change of total energy, which figures on the left-hand side
of the energy balance, equals the total dissipation rate.   
However, in the more general case the rate of change of the total energy 
is given by the difference of energy input and dissipation. That is, 
in the most general case the total energy dissipation rate is given by
\beq
\vep(t) =\iota(t) -d_t E (t)\ .
\eeq  
For the stationary case $d_t E (t)=0$ and one obtains $\vep(t)=\iota(t)$.
For the freely decaying case $\iota(t) = 0$ and the change in total energy is due to 
dissipation only, that is $-d_t E (t)=\vep(t)$. 
In terms of Els\"asser variables $\vep(t)$ can also be expressed as
\beq
\vep(t) = \iota(t)-d_t E(t) = \iota(t) - d_tE^\pm(t) \mp d_tH_c(t) \ ,
\eeq
where $E^\pm(t)$ denote the Els\"asser energies. Since we have related the total dissipation rate 
to the rate of change of the Els\"asser energies, we are now in a position to consider the energy 
balance equations for $\vec{z}^\pm$, which are
stated here for the most general case of homogeneous forced nonstationary MHD flows without a mean 
magnetic field
\begin{align}
-\partial_t E^\pm(t) +I^\pm(r,t)  
 = & -\frac{3}{4} \partial_t B_{LL}^{\pm\pm}(r,t) - \frac{\partial_r}{r^4}
\left(\frac{3r^4}{2}C^{\pm\mp\pm}_{LL,L}(r,t) \right) \nonumber \\
& \ \ \ +\frac{3}{4r^4} \partial_r \left(r^4 \partial_r(\nu+\mu)B^{\pm}_{LL}(r,t) \right) \nonumber \\
& \ \ \ +\frac{3}{4r^4} \partial_r \left(r^4 \partial_r(\nu-\mu)B^{\mp}_{LL}(r,t) \right) \ ,
\label{eq:elsasser_balance}
\end{align}
where $I^\pm(r,t)$ are (scale-dependent) energy input terms and 
\begin{align}
C^{\pm\mp\mp}_{LL,L}(r,t)&= \langle z_L^\pm(\vec{x},t) z_L^\mp(\vec{x},t) z_L^\pm(\vec{x} + \vec{r},t) \rangle \ ,  \\
B_{LL}^{\pm\pm}(r,t)&= \langle (\delta z_L^\pm(\vec{r},t))^2 \rangle \ , \\ 
B_{LL}^{\pm\mp}(r,t)&= \langle \delta z_L^\pm(\vec{r},t) \delta z_L^\mp(\vec{r},t) \rangle \ , 
\end{align}
are the third-order longitudinal correlation function and the
second-order structure functions of the Els\"{a}sser fields, respectively. 
As can be seen from the definition, the third-order correlation function
scales with $(z^\pm)^2z^\mp$, where $z^\pm$ denote the respective 
rms values of the Els\"asser fields. This permits a consistent nondimensionalization
of the Els\"asser vKHE using the appropriate quantities 
defined in terms of Els\"asser variables. As such the complication that 
arose if the energy balance was written in terms of $\vec{b}$ and $\vec{u}$
can be circumvented.  
This motivates the definition of the dimensionless Els\"asser dissipation rates 
as
\beq
\Ceps^\pm(t) \equiv \frac{\vep(t) L_{\pm}(t)}{z^\pm(t)^2 z^\mp(t)} \ ,
\label{eq:cepspm_defn}
\eeq
where 
\beq
L_{\pm}(t)=\frac{3\pi}{8E^\pm(t)} \int_\Omega d \vk  \ k^{-1} \langle |\vec{z}^\pm(\vk,t)|^2 \rangle \ ,
\eeq
are the integral scales defined with respect to $\vec{z}^\pm$
\footnote{The scaling is ill-defined for the 
(measure zero) cases $\vec{u} = \pm \vec{b}$,
which correspond to exact solutions to the MHD equations where
the nonlinear terms vanish. Thus no turbulent transfer is possible,
and these cases are not amenable to an analysis
which assumes nonzero energy transfer \cite{Politano98}.} 
.
For balanced MHD turbulence, {\em i.e.}~$H_c=0$, one should expect $\Ceps^+(t) = \Ceps^-(t)$, since
\beq
E^\pm(t) = 2E(t)\pm 2H_c(t) = 2E(t)  \ .
\eeq
Therefore all quantities defined with respect to the rms fields $z^+$ and
$z^-$ should be the same in this case.
Finally, the dimensionless dissipation rate $\Ceps(t)$ is defined as
\beq
\Ceps(t) = \Ceps^+(t) + \Ceps^-(t) \equiv  \frac{\vep(t) L_+(t)}{{z^+(t)}^2 z^-(t)} + \frac{\vep(t) L_-(t)}{{z^-(t)}^2 z^+(t)} \ .
\label{eq:ceps_defn}
\eeq
Using the definition given in Eq.~\eqref{eq:cepspm_defn},
the Els\"asser energy balance equations \eqref{eq:elsasser_balance}
can now be consistently nondimensionalized. 
For conciseness the explicit time and spatial dependences 
are from now on omitted, unless there is a particular point to make. 

\subsection{Dimensionless von K\'arm\'an-Howarth equations}
By introducing the nondimensional variables $\sigma_\pm=r/L_{\pm}$ \cite{Wan12} and
 non-dimensionalising Eq.~\eqref{eq:elsasser_balance} 
as proposed in the definitions of $\Ceps^\pm$ given in Eq.~\eqref{eq:cepspm_defn} one obtains
\begin{align}
\label{eq:evol_z_nondim}
-\left(d_t E^\pm -I^\pm \right) 
\frac{L_\pm}{{z^\pm}^2 z^\mp} 
=
& -\frac{1}{\sigma_\pm^4} \partial_{\sigma_\pm} \left(\frac{3\sigma_\pm^4C^{\pm\mp\pm}_{LL,L}}{2{z^\pm}^2 z^\mp}\right)  
 -\frac{L_{z^\pm}}{{z^\pm}^2z^\mp}
\partial_t \frac{3B_{LL}^{\pm\pm}}{4} 
\nonumber \\
& +\frac{\mu + \nu}{L_\pm z^\mp} \frac{3}{4\sigma_\pm^4} 
\left(\sigma_\pm^4 \partial_{\sigma_\pm}\frac{B_{LL}^{\pm\pm}}{{z^\pm}^2} \right) \nonumber \\
& +\frac{\nu-\mu}{L_\pm z^\pm} \frac{3}{4\sigma_\pm^4} 
\left(\sigma_\pm^4 \partial_{\sigma_\pm}\frac{B_{LL}^{\pm\mp}}{{z^\pm}z^\mp} \right) \ .
\end{align}
Before proceeding further, the scale-dependent forcing term on the left-hand side
of this equation needs to be analyzed in some detail in order to clarify
its relation to the energy input rates $\iota$ and $\iota^\pm$. 
The Els\"asser energy input $I^\pm$ is given by 
\beq
I^\pm(r) = \frac{3}{r^3}\int_0^r d r' r'^2 \langle \vec{z}^{\pm}(\vx + \vr') \cdot \vec{f}^{\pm}(\vx) \rangle \ .
\eeq
\blue{Since the energy input rate is given by
$\iota^\pm = \langle \vec{z}^{\pm}(\vx) \cdot \vec{f}^{\pm}(\vx) \rangle$,
the correlation function 
can be expressed as
\beq
\langle \vec{z}^{\pm}(\vx + \vr) \cdot \vec{f}^{\pm}(\vx) \rangle = \iota^\pm \phi^\pm(r/L_f) \ ,
\eeq
where $\phi^\pm$ are dimensionless even functions of $r/L_f$ satisfying $\phi^\pm(0) = 1$  
and $L_f$ the characteristic scale of the forcing. 
At scales much smaller than the forcing scale, i.e. for $r/L_f << 1$, for suitable types of
 forces 
$\phi^\pm(r/L_f)$ can be expanded in a Taylor series 
\cite{Novikov65}, leading to the following expression for the energy input }
\beq
\label{eq:input_expansion}
I^\pm(r) = \frac{3}{r^3}\int_0^r d r' r'^2 \iota^\pm 
\left[ 1+ \left(\frac{r}{L_f}\right)^2 \frac{\partial^2 \phi^\pm}{2\partial(r/L_f)^2}\Big|_{r/L_f = 0} 
+ O\left(\left(\frac{r}{L_f}\right)^4\right) 
\right] \ .
\eeq
In the limit of infinite Reynolds number the inertial range 
extends through all wavenumbers, formally implying that 
\blue{$L_f \to \infty$, 
where Eq.~\eqref{eq:input_expansion} implies $I^\pm(r)\to \iota^\pm$}. 
Therefore it should be possible 
to split the term $I^\pm(r)$ into a constant, $\iota^\pm$,
and a scale-dependent term $J^\pm(r)$, which encodes the additional scale dependence introduced by 
realistic, finite Reynolds number forcing. For consistency, 
this scale-dependent term must vanish in the formal limit $Re \to \infty$. 
This can be achieved by writing $I^\pm(r)$ in terms of the 
correlation of the force and Els\"asser field increments 
\beq
I^\pm(r) = \iota^\pm - \frac{3}{2r^3} \int_0^r d r' r'^2 \langle \delta \vec{z}^{\pm} \cdot \delta \vec{f}^{\pm} \rangle \ .
\eeq
Therefore we define 
\beq
J^\pm(r) = -\frac{3}{2r^3}\int_0^r d r' r'^2 \langle \delta \vec{z}^{\pm} \cdot \delta \vec{f}^{\pm} \rangle \ ,
\eeq
where $\lim_{Re \to \infty} J^\pm(r) = 0$. Hence the energy input  
$I^\pm(r)$ can be expressed as the sum of the scale-independent energy input rate $\iota^\pm$ and 
a scale-dependent term which vanishes in the formal limit $Re \to \infty$
\beq
\label{eq:input}
I^\pm(r) = \iota^\pm + J^\pm(r) \ ,
\eeq
with $\lim_{Re \to \infty} J^\pm(r) = 0$. 
Substitution of Eq.~\eqref{eq:input} into the nondimensionalized energy balance 
Eq.~\eqref{eq:evol_z_nondim} leads to the dimensionless version of the Els\"asser
vKHE for homogeneous MHD turbulence
in the most general case for nonstationary flows at any magnetic Prandtl number
\begin{align}
\Ceps^\pm
= & -\frac{\partial_{\sigma_\pm}}{\sigma_\pm^4} \left(\frac{3\sigma_\pm^4C^{\pm\mp\pm}_{LL,L}}{2{z^\pm}^2 z^\mp}\right) 
+\frac{L_\pm}{{z^\pm}^2 z^\mp}\left( \pm d_t H_{c} -\partial_t \frac{3B_{LL}^{\pm\pm}}{4} -J^\pm \mp \iota_{H_c} \right ) \nonumber \\ 
& +\frac{1}{R_\mp} \frac{3\partial_{\sigma_\pm}}{2\sigma_\pm^4} \left(\sigma_\pm^4\partial_{\sigma_\pm}\frac{B_{LL}^{\pm\pm}}{{z^\pm}^2} \right) 
+\frac{1}{R'_\pm} \frac{3\partial_{\sigma_\pm}}{2\sigma_\pm^4} \left(\sigma_\pm^4\partial_{\sigma_\pm}\frac{B_{LL}^{\pm\mp}}{{z^\pm}z^\mp} \right) \ ,
\label{eq:evol_z_scaled_long}
\end{align}
where $R_{\mp}$ and $R'_{\pm}$ denote generalized large-scale Reynolds numbers given by 
\begin{align}
R_{\mp}&=z^\mp L_{\pm}/(\nu+\mu) \ \ \mbox{and} \ \  R'_{\pm}=z^\pm L_{\pm}/(\nu-\mu) \ . 
\label{eq:gen_Rey}
\end{align}  
In order to express Eq.~\eqref{eq:evol_z_scaled_long} more concisely, the following dimensionless functions are defined 
\begin{align}
g^{\pm\mp\pm} &= \frac{C^{\pm\mp\pm}_{LL,L}}{{z^\pm}^2z^\mp} \ , \\
h^{\pm\pm} &= \frac{B_{LL}^{\pm\pm}}{{z^\pm}^2} \ , \\
h^{\pm\mp} &= \frac{B_{LL}^{\pm\mp}}{{z^\pm}z^\mp} \ , \\
H^{\pm\pm}  &=\frac{L_\pm}{{z^\pm}^2z^\mp} \partial_t B_{LL}^{\pm\pm} \ , \\ 
F^\pm  &=\frac{L_\pm}{{z^\pm}^2z^\mp} J^\pm\ , \\
G^\pm & = \frac{L_\pm}{{z^\pm}^2z^\mp} d_t H_c \ , \\ 
Q^\pm & = \frac{L_\pm}{{z^\pm}^2z^\mp} \iota_{H_c} \ , 
\end{align}
such that Eq.~\eqref{eq:evol_z_scaled_long} can be written as
\begin{align}
\Ceps^\pm = 
& -\frac{\partial_{\sigma_\pm}}{\sigma_\pm^4} \left(\frac{3\sigma_\pm^4}{2}g^{\pm\mp\pm}\right) 
\pm G^\pm -\frac{3}{4}H^{\pm\pm} -F^\pm \mp Q^\pm \nonumber \\
& +\frac{3}{R_\mp} \frac{\partial_{\sigma_\pm}}{\sigma_\pm^4} \left(\sigma_\pm^4 \partial_{\sigma_\pm}h^{\pm\pm} \right)
+\frac{3}{R'_\mp} \frac{\partial_{\sigma_\pm}}{\sigma_\pm^4} \left(\sigma_\pm^4 \partial_{\sigma_\pm}h^{\pm\mp} \right) \ .
\label{eq:evol_z_scaled}
\end{align}
This equation can be applied to the two simpler cases of freely decaying and stationary MHD turbulence
by setting the corresponding terms to zero. For the case of free decay there are no external forces 
therefore $F^\pm = 0$, while for the stationary case the terms $G^\pm$ and $H^\pm$ vanish. 
A further simplification concerns the case $Pm =1$, that is $\nu=\mu$, where the inverse of the 
generalized Reynolds numbers $R'_\pm$ vanish. 
In this case the evolution of $\Ceps^\pm$ depends only on $R_\mp$, and an approximate analysis 
using asymptotic series is possible. Most numerical results are concerned 
with this case due to computational constraints, hence it would be very difficult to 
test an approximate equation against DNS data if not only $Re$ but also $Pm$ needs to be varied.
From now on the magnetic Prandtl number is therefore set to unity, keeping in mind that the analysis
could be extended to $Pm \neq 1$ provided the approximate equation derived in the following section
is consistent with DNS data.

\subsection{Asymptotic analysis for the case $Pm=1$} 
Equation \eqref{eq:evol_z_scaled} suggests a dependence of $\Ceps^\pm$ on $1/R_{\mp}$, 
however, the structure and correlation functions also have a dependence on
Reynolds number, which describes their deviation from their respective inertial-range forms. 
The highest derivative in Eq.~\eqref{eq:evol_z_scaled} is multiplied by the 
small parameter $1/R_\mp$, which suggests
that this equation may be viewed as singular perturbation problem amenable 
to asymptotic analysis \cite{Lundgren02}. 
The Els\"asser vKHE was rescaled by 
the rms values of the Els\"asser fields and the corresponding integral length scales, where
the integral scales are by definition the large-scale quantities, 
the interpretation in hydrodynamics usually being that they 
represent the size of the largest eddies.  
As such, the nondimensionalization was carried out with respect 
to quantities describing the large scales, that is,
with respect to `outer' variables. Hence outer asymptotic 
expansions of the nondimensional structure and correlation functions are 
considered with respect
to the inverse of the (large-scale) generalized Reynolds numbers $1/R_{\mp}$. 
We point out that 
the case $Pm \neq 1$ would require expansions in two parameters, 
where the cases $Pm >1$ and $Pm <1$ must be treated separately due to a sign change in $R'_\pm$
between the two cases. 
    
The formal asymptotic series of a generic function $f$ [used for conciseness in place of
the functions on the right-hand side of Eq.~\eqref{eq:evol_z_scaled}]
up to second order in $1/R_\mp$ reads
\begin{equation}
\label{eq:asymp_F}
f= f_{0}+ \frac{1}{R_\mp}f_{1} 
+ \frac{1}{R_\mp^2}f_{2} + O(R_\mp^{-3}) \ .  
\end{equation}
After substitution of the expansions into
Eq.~\eqref{eq:evol_z_scaled}, collecting terms of the same order in $1/R_\mp$,
one arrives at equations describing the behavior of $\Ceps^+$ and $\Ceps^-$
\beq
\Ceps^\pm = \Cinf^\pm + \frac{C^\pm}{R_{\mp}} + \frac{D^\pm}{R_{\mp}^2} + O(R_{\mp}^{-3}) \ ,
 \label{eq:model+}
\eeq
up to second order in $1/R_{\mp}$,
using the coefficients $\Cinf^\pm$, $C^\pm$ and $D^\pm$ defined as
\begin{align}
\label{eq:cinf+}
\Cinf^\pm &= -\frac{\partial_{\sigmapm}}{\sigmapm^4} \left(\frac{3\sigmapm^4}{2}g_0^{\pm\mp\pm}\right)
  \pm  G^\pm - \frac{3}{4} H_{0}^{\pm\pm} \mp Q^\pm \ , \\
\label{eq:c+}
C^\pm &=\frac{3\partial_{\sigmapm}}{\sigmapm^4} \left[ \sigmapm^4 \left( \partial_{\sigmapm} 
h_{0}^{\pm\pm} - \frac{g_1^{\pm\mp\pm}}{2} \right)\right] 
 \mp F_1^\pm - \frac{3}{4} H_1^{\pm\pm}\ , \\ 
\label{eq:d+}
D^\pm &=\frac{3\partial_{\sigmapm}}{\sigmapm^4} \left[ \sigmapm^4 \left( \partial_{\sigmapm} 
h_{1}^{\pm\pm} - \frac{g_2^{\pm\mp\pm}}{2} \right)\right] 
 \mp F_2^\pm - \frac{3}{4} H_2^{\pm\pm}\ , 
\end{align}
in order to write Eq.~\eqref{eq:evol_z_scaled} in a more concise way.
The zero-order term in the expansion of the function $F^\pm$ vanishes, 
since $F^\pm$ corresponds to the scale-dependent part $J^\pm$ of the energy
input which vanishes in the limit $R_\mp \to \infty$, hence $F_0^\pm =0$.
According to the definition of $\Ceps$ in Eq.~\eqref{eq:ceps_defn}, the asymptote $\Cinf$ is given by 
\beq
\Cinf = \Cinf^+ + \Cinf^- \ ,
\eeq
and using the definition of the generalized Reynolds numbers, which implies
$R_+ = (L_-/L_+)(z^+/z^-)R_-$ one can define
\beq
C=  C^+ + \frac{L_-}{L_+} \frac{z^+}{z^-} C^-  \ ,
\eeq
($D$ is defined analogously), resulting in the following expression for the dimensionless
dissipation rate 
\beq
\Ceps = \Cinf + \frac{C}{R_-} + \frac{D}{R_-^2} + O(R_-^{-3}) \ .
 \label{eq:model}
\eeq
Since the time dependence of the various quantities in this problem has been
suppressed for conciseness, it has to be emphasized 
that Eq.~\eqref{eq:model} is time dependent, including the Reynolds
number $R_-$. 
Equation \eqref{eq:model} in conjunction with eqs.~\eqref{eq:cinf+}-\eqref{eq:d+}
is the most general asymptotic expression for the Reynolds number dependence 
of $\Ceps$ developed so far. It is applicable for freely decaying, stationary and non-stationary 
MHD turbulence in the presence of external forces, and it may be applied to the corresponding 
problem in non-conducting fluids 
by setting $\vec{b} =0$. As such it extends previous results for freely decaying 
MHD turbulence \cite{Linkmann15a}, as well as for the stationary case in 
homogeneous isotropic turbulence of non-conducting fluids \cite{McComb15a}. 

For nonstationary MHD turbulence at the peak of dissipation the term $H_0^{\pm\pm}$ in Eq.~\eqref{eq:cinf+}
vanishes for constant flux of cross-helicity
(that is, $d_t^2 H_c = 0$),
since in the infinite Reynolds number limit the
second-order structure function will have its inertial range form
at all scales. By self-similarity the spatial and temporal dependences 
of e.g.~$B_{LL}^{++}$
should be separable in the inertial range, that is 
\beq
B_{LL}^{++}(r,t) \sim (\vep^+(t)r)^{\alpha} \ ,
\eeq
for some value $\alpha$, and
\beq
\partial_t B_{LL}^{++} \sim \alpha\vep^+(t)^{\alpha-1} \ d_t \vep^+ r^\alpha \ .
\eeq
At the peak of dissipation 
\beq
d_t \vep^+|_{t_{peak}} = 
d_t\vep|_{t_{peak}} -d_t^2 H_c = d_t\vep|_{t_{peak}} =0 \ ,
\eeq
which implies $H_0^{++}(t_{peak})=0$.
Equation \eqref{eq:cinf+} taken for nonstationary flows at the peak of 
dissipation is thus identical to Eq.~\eqref{eq:cinf+} for stationary flows,
which suggests that at this point in time a nonstationary flow 
may behave similarly to a stationary flow. We will come back to this point 
in Sec.~\ref{sec:numerics}. 
Due to selective decay, that is the faster decay of the total energy
compared to $H_c$ and $H_m$ \cite{Biskamp93}, in most situations one could expect $d_t H_c$ to
be small compared to $\vep$ in the infinite Reynolds number limit.
In this case $G^{\pm} \simeq 0$ and
\beq
\Cinf^\pm(t_{peak}) = -\frac{\partial_{\sigmapm}}{\sigmapm^4} \left(\frac{3\sigmapm^4}{2}g_0^{\pm\mp\pm} \right) \ ,
\eeq
which recovers the inertial-range scaling results of Ref.~\cite{Politano98} and reduces
to Kolmogorov's four-fifth law for $\vec{b}=0$.

\subsection{Relation of $\Cinf$ to energy and cross-helicity fluxes}
In analogy to hydrodynamics, the asymptotes $\Cinf^\pm$ should 
describe the total energy flux, that is  
the contribution of the cross-helicity flux to the
Els\"{a}sser flux should be canceled by the respective terms $G^\pm$ and $Q^\pm$ in  
Eq.~\eqref{eq:cinf+}.
However, since this is not immediately obvious from the derivation, 
further details are given here. 
For 
nonstationary turbulence at the peak of dissipation, Eq.~\eqref{eq:cinf+} for the 
asymptotes $\Cinf^\pm$ reduces to
\beq
\label{eq:cinf+_tp}
\Cinf^\pm = -\frac{\partial_{\sigmapm}}{\sigmapm^4} \left(\frac{3\sigmapm^4}{2}g_0^{\pm\mp\pm}\right)  \pm  G^\pm \mp  Q^\pm \ .
\eeq
The dimensional version of this equation is
\beq
\label{eq:cinf+_tp_dim}
\vep = -\frac{\partial_r}{r^4} \left(\frac{3r^4}{2}C_{LL,L}^{\pm\mp\pm}\right)  \pm  d_t H_c \mp \iota_{H_c}\ ,
\eeq
where it is assumed that the function $C_{LL,L}^{\pm\mp\pm}$ has its inertial range form corresponding to $g_0^{\pm\mp\pm}$.
The function $C_{LL,L}^{\pm\mp\pm}$ can also be expressed through the Els\"asser increments \cite{Politano98}  
\beq
C_{LL,L}^{\pm\mp\pm}
=\frac{1}{4}\left ( \langle (\delta  z_L^\pm(\vr))^2 \delta z_L^\mp(\vr)\rangle
-2\langle z_L^\pm(\vec{x}) z_L^\pm(\vec{x}) z_L^\mp(\vec{x} + \vec{r}) \rangle \right ) \ ,
\eeq
which can be written in terms of the primary fields $\vec{u}$ and $\vec{b}$ as
\begin{align}
\label{eq:CLL_ub}
C_{LL,L}^{\pm\mp\pm}&=
 \frac{1}{4}\frac{2}{3}\langle(\delta u_L(\vr))^3 - 6  b_L(\vec{x})^2 u_L(\vec{x}+\vec{r}) \rangle \nonumber \\
& \ \ \mp \frac{1}{4} \frac{2}{3}\langle(\delta b_L(\vr))^3 - 6  u_L(\vec{x})^2 b_L(\vec{x}+\vec{r}) \rangle \ ,
\end{align}
(see e.g.~Ref.~\cite{Politano98}).
The two terms on the first line of Eq.~\eqref{eq:CLL_ub} are the 
flux terms in the evolution equation of the total energy, while
the two terms on last line correspond to the flux
terms in the evolution equation of the cross-helicity \cite{Politano98}.
Now Eq.~\eqref{eq:cinf+_tp_dim} can be expressed in terms of the primary fields 
\begin{align}
\label{eq:cinf+_tp2}
\vep &= -\frac{\partial_r}{r^4} \left(\frac{3r^4}{2}C_{LL,L}^{\pm\mp\pm}\right)  \pm  d_t H_c \mp \iota_{H_c} \nonumber \\
&=-\frac{\partial_r}{r^4} \left(\frac{r^4}{4}  \langle(\delta u_L(\vr))^3 - 6  b_L(\vec{x})^2 u_L(\vec{x}+\vec{r}) \rangle \right) \nonumber \\
& \ \  \ \pm \frac{\partial_r}{r^4} \left(\frac{r^4}{4}  \langle(\delta b_L(\vr))^3 - 6  u_L(\vec{x})^2 b_L(\vec{x}+\vec{r}) \rangle  \right) 
\pm d_t H_c \mp \iota_{H_c} \nonumber \\
&= \vep_T \pm \vep_{H_c} \pm d_t H_c \mp \iota_{H_c}= \vep_T \ ,
\end{align}
where $\vep_T$ is the flux of total energy and $\vep_{H_c}$
the cross-helicity flux, which must equal $-d_t H_c + \iota_{H_c}$ 
for nonstationary MHD turbulence. Thus the
contribution from the third-order correlator $C_{LL,L}^{\pm\mp\pm}$
resulting in $\vep_{H_c}$ is canceled by $d_t H_c - \iota_{H_c}$, or,
after nondimensionalization, the cross-helicity flux
$\vep_{H_c} L_{\pm}/[(z^\pm)^2z^\mp]$ is canceled by $G^\pm-Q^\pm$.
The two simpler cases of freely decaying and stationary MHD turbulence are recovered by setting either $Q^\pm =0$ (free decay) or 
$G^{\pm} =0$ (stationary case).

\subsection{Nonuniversality} \label{sec:nonuniversality}
Since $\Cinf$ is a measure of the flux of total energy across different scales in
the inertial range, differences for the value of this asymptote should be
expected for systems with different initial values for the ideal invariants
$H_m$ and $H_c$.
The flux of total energy and thus the asymptote $\Cinf$ is an averaged quantity. This
implies that cancellations between forward and inverse fluxes may take place leading on average 
to a positive value of the flux, that is, forward transfer from the large scales to the small scales.  
In case of $H_m \neq 0$, the value of $\Cinf$ should
therefore be {\em less} than for $H_m=0$ due to a more pronounced inverse
energy transfer in the helical case, the result of which is {\em less} average forward transfer and
thus a smaller value of the (average) flux of total energy.
For $H_c \neq 0$ the asymptote $\Cinf$ is expected to be smaller than for $H_c =0$,
since alignment of $\vec{u}$ and $\vec{b}$ weakens the coupling
of the two fields in the induction equation, leading to less transfer
of magnetic energy across different scales and presumably also less transfer
of kinetic to magnetic energy.

Furthermore, from an analysis of helical triadic interactions in ideal MHD carried out in 
Ref.~\cite{Linkmann16a} it may be expected that 
high values of cross-helicity have a different effect on the asymptote $\Cinf$, 
depending on the level of magnetic helicity. The analytical results suggested that 
the cross-helicity may have an asymmetric effect on the nonlinear transfers in the sense
that the self-ordering inverse triadic transfers are less quenched by high levels of
$H_c$ compared to the forward transfers. The triads contributing to 
inverse transfers were mainly those where magnetic field modes of like-signed helicity interact, 
and so for simulations with maximal initial magnetic helicity the dynamics will be dominated
by these triads. If the inverse fluxes are less affected by the cross-helicity than the forward
fluxes, then the expectation is that for a comparison of the value of $\Cinf$ between 
systems with (i) high $H_m$ and $H_c$, (ii) high $H_m$ and $H_c=0$, (iii) $H_m=0$ and high $H_c$ and 
finally (iv) $H_m=0$ and $H_c=0$, the value of $\Cinf$ should diminish more between cases (i) and (ii) compared to 
between cases (iii) and (iv). Such a comparison is carried out in Sec.~\ref{sec:numerics} using DNS data.
 
As can be seen from Eqs.~\eqref{eq:cinf+}-\eqref{eq:d+},  
the force does not explicitly enter in the asymptote $\Cinf$ but
does so in the coefficients $C$ and $D$. 
Therefore a dependence of $C$ and $D$, and hence of the 
approach to the asymptote, on the force may be expected, while $\Cinf$ 
appears to be unaffected by the external force.   
However, different external forces will lead to different
energy transfer scenarios, e.g. mainly dynamo and inertial transfer or mainly 
conversion of magnetic to kinetic energy due to a strong Lorentz force, 
therefore the asymptote will be implicitly influenced by that. 
In short, nonuniversal values of $\Cinf$, $C$ and $D$ are expected depending 
on the level of the ideal invariants and the type of external force.
We will address this point in further detail in Secs.~\ref{sec:numerics} and \ref{sec:conclusions}. 

\section{Comparison to DNS data}
\label{sec:numerics}
Before comparing Eq.~\eqref{eq:model} with DNS data 
the numerical method is briefly outlined.
Equations \eqref{eq:momentum}-\eqref{eq:incompr} are solved numerically
in a three-dimensional periodic domain of length $L_{box}=2 \pi$
using a fully de-aliased pseudospectral MHD code \cite{Berera14a,Linkmann_thesis}.
Both the initial magnetic
and velocity fields are random Gaussian with zero mean with energy spectra 
given by
\beq
E_{mag,kin}(k)=A k^4\exp(-k^2/(2k_0)^2) \ ,
\eeq
where $A \geqslant 0$ is a real number which can be adjusted according to
the desired amount of initial energy. The wavenumber $k_0$ which locates
the peak of the initial spectrum is taken to be $k_0 = 5$ unless otherwise stated. 
No background magnetic field is imposed.

Several series of simulations have been carried out for stationary and freely decaying MHD 
turbulence. In the case of free decay the dependence of the asymptote on the initial level
of the ideal invariants is studied. For the stationary simulations all helicities are 
initially negligible while the influence of different forcing methods is assessed by applying 
three different external mechanical forces labeled $\vec{f}_1$, $\vec{f}_2$ and $\vec{f}_3$
to maintain the simulations in stationary state, resulting in three different series of stationary 
DNSs. The forces always act at wavenumbers $k \leqslant k_f =2.5$, {\em i.e.~}at the large scales.
The first type of mechanical force $\vec{f}_1$
corresponds to the DNS series ND in Tbl.~\ref{tbl:simulations_forced} and is 
given by 
\begin{align}
 \hat{\vec{f}_1}(\vec{k},t) &=
      (\iota_{kin}/2 E_f) \hat{\vec{u}}(\vec{k},t) \quad
\text{for} \quad  0 < \lvert\vec{k}\rvert < k_f ; \nonumber \\
  &= 0   \quad \textrm{otherwise},
\label{eq:forcing1}
\end{align}
where $\hat{\vec{f}_1}(\vec{k},t) $ is the Fourier transform of the forcing 
and $E_f$ is the total energy contained in the forcing
band. 
The second type of mechanical force $\vec{f}_2$, which corresponds 
to the DNS series HF in Tbl.~\ref{tbl:simulations_forced} is a random $\delta(t)$-correlated 
process. It is based on a decomposition 
of the Fourier transform of the force into helical modes and 
has the advantage that the helicity of the force can be adjusted 
at each wavevector \cite{Brandenburg01}, which gives 
optimal control over the helicity injection.   
For all simulations using this type of forcing the relative 
helicity of the force was set to zero. 
The third type of mechanical force \cite{Dallas14a} corresponds 
to the DNS series SF in Tbl.~\ref{tbl:simulations_forced} and is given by
\beq
\vec{f}_3 = f_0 \sum_{k_f}
 \begin{pmatrix}
    \sin{k_fz} + \sin{k_fy} \\
    \sin{k_fx} + \sin{k_fz} \\
    \sin{k_fy} + \sin{k_fx} 
  \end{pmatrix} \ ,
\label{eq:forcing2}
\eeq
where $f_0$ is an adjustable constant. This type of force is nonhelical by construction. 

All three forces have been used in several simulations of stationary homogeneous MHD turbulence. 
The scheme labeled $\vec{f}_1$ was shown by Sahoo \etal~\cite{Sahoo11} to keep the helicities at negligible levels
even though zero helicity injection cannot be guaranteed with this forcing scheme. 
At very low Reynolds number this conservation of helicities appears to be broken and 
induces peculiar self-ordering effects \cite{McComb15b,Linkmann15b}. 
The adjustable helicity forcing 
$\vec{f}_2$ has been extensively used in the literature \cite{Brandenburg01,Mueller12,Malapaka13,Linkmann16b}, mainly when nonzero 
levels of kinetic \cite{Brandenburg01} or magnetic \cite{Mueller12,Malapaka13,Linkmann16b} helicity injection are required. The 
third forcing scheme $\vec{f}_3$ has been employed in the simulations by Dallas and Alexakis \cite{Dallas14a}, where 
it was shown that despite zero injection of all helicities, the system self-organized into large-scale fully helical 
states as soon as electromagnetic forces were applied. 
 \begin{table}[]
 \begin{center}
 \begin{tabular}{cccccccccccccccc}
   Run id & $N^3$  & $k_{max}\eta_{mag}$ & $k_{max}\eta_{kin}$& $R_{-}$ & $R_L$ & $\Rl$ & $\vep_{mag}/\vep$ & $\vep_{kin}/\vep$ & $\mu=\nu $  & $k_{0}$ & \#  & $\Ceps$
  & $\sigma$ & $\rho_c(0)$ \\
  \hline
   H1 & $128^3$ & 1.30 & 1.61 & 33.37 & 25.28 & 14.87 & 0.70 & 0.30 & $0.009$ & 5 & 10 & 0.756 & 0.008 & 0\\
  H2 & $256^3$ & 2.42 & 3.00 & 37.77 & 27.81 & 15.85 & 0.70 & 0.30 & $0.008$ &  5 & 10 & 0.704  & 0.007 & 0\\
  H3 & $512^3$ & 1.38 & 1.70  & 50.81 & 35.08 & 18.55 & 0.70 & 0.30 & $0.002$  & 15 & 10  & 0.608 & 0.001 & 0 \\
  H4 & $256^3$ & 1.80 & 2.23 & 61.14 & 40.63 & 20.34 & 0.70 & 0.30 & 0.005  &  5 & 10 & 0.569 & 0.006 & 0 \\
  H5 & $256^3$ & 1.59 & 1.92 & 76.72 & 48.73 & 23.11 & 0.68 & 0.32 & 0.004 &  5 & 10 & 0.510 & 0.005 & 0\\
  H6 & $1024^3$ & 1.38 & 1.53 & 89.32 &55.51 & 25.76 & 0.60 & 0.40 & $0.00075$  & 23 & 10 & 0.4589 & 0.0003 & 0\\
  H7 & $256^3$ & 1.29 & 1.57 & 102.53 & 60.65 & 26.91 & 0.69& 0.31 & $0.003$ & 5 & 10 & 0.450 & 0.004 & 0\\
  H8 & $512^3$ & 2.33 & 2.78 & 123.17 & 69.40 & 29.67 & 0.67 & 0.33 & $0.0025$ & 5 & 10 & 0.419 & 0.003 & 0\\
  H9 & $512^3$ & 2.01 & 2.40 & 154.67 & 83.06 & 33.84 & 0.67 & 0.33& $0.002$ & 5 & 10 & 0.384 & 0.003 & 0\\
  H10 & $512^3$ & 1.45 & 1.66 & 255.89 & 123.97 & 45.21 & 0.63 & 0.37 & $0.0012$ & 5 & 10 & 0.320 & 0.004 &0\\
  H11 & $528^3$ & 1.31 & 1.51 & 308.69 & 143.71 & 50.18 & 0.64  & 0.36& $0.001$ & 5 & 10 & 0.310 & 0.004& 0\\
  H12 & $1024^3$ & 2.03 & 2.28 & 441.25 & 194.38 & 61.39 & 0.61 & 0.39& $0.0007$  & 5 & 5 & 0.281 & 0.002 & 0\\
  H13 & $1032^3$ & 1.38 & 1.53 & 771.34 & 309.08 & 82.97 & 0.60 & 0.40& $0.0004$  & 5 & 5 & 0.268 & 0.001 & 0\\
  H14 & $1024^3$ & 1.24 & 1.38 & 885.05 & 358.72 & 88.76 & 0.61 & 0.39& $0.00035$  & 5 & 5 & 0.265 & 0.002 & 0\\
  H15 & $2048^3$ & 1.35 & 1.48 & 2042.52 & 724.71 & 136.25 & 0.59& 0.41& $0.00015$  & 5 & 1 & 0.250 & - & 0\\
  \hline
  CH06H1 & $512^3$ & 2.17  & 2.50 & 124.89 & 108.81 & 49.88 & 0.64 & 0.36 & $0.002$ & 5 & 1 & 0.380 & - & 0.6 \\
  CH06H2 & $512^3$ & 1.57 & 1.78 & 207.61 & 171.87 & 68.57 & 0.62 & 0.38& $0.0012$ & 5 & 5 & 0.309 & 0.002 & 0.6 \\
  CH06H3 & $1024^3$ & 2.21 & 2.44 & 351.52 & 277.21 & 95.31 & 0.60 & 0.40 & $0.0007$ & 5 & 1 & 0.260 & - & 0.6 \\
  CH06H4 & $1024^3$ & 1.76 & 1.93 & 491.50 & 380.70 & 116.85 & 0.59& 0.41 & $0.0005$ & 5 & 1 & 0.236 & - & 0.6 \\
  CH06H5 & $1024^3$ & 1.37 & 1.50 & 696.19 & 523.08 & 132.48 & 0.59 & 0.41& $0.00035$ & 5 & 1 & 0.231 & - & 0.6 \\
  \hline
  \end{tabular}
  \end{center}
 \caption{Specifications of decaying simulations with maximal initial magnetic helicity \cite{data15a,Linkmann15a}.
 $R_L$ denotes the integral-scale Reynolds number,
$R_{\lambda}$ the Taylor-scale Reynolds number,
 $R_{-}$ the generalized Reynolds number, $\mu$ the
 magnetic resistivity, $k_0$ the peak wavenumber of the initial energy spectra,
 $k_{max}$ the largest resolved wavenumber, 
\blue{$\eta_{mag} = (\mu^3/\vep_{mag})^{1/4} $ and $\eta_{kin} = (\nu^3/\vep_{kin})^{1/4}$ the magnetic
and kinetic Kolmogorov microscales, respectively, at the peak of total dissipation},
 \# the ensemble size, $\Ceps$ the dimensionless total dissipation rate, $\sigma$ the
 standard error on $\Ceps$ and $\rho_c(0)$ the initial relative cross helicity. All Reynolds numbers are measured
 at the peak of total dissipation.
 }
 \label{tbl:simulations_h}
 \end{table}

 \begin{table}[]
 \begin{center}
 \begin{tabular}{cccccccccccccccc}
   Run id & $N^3$  & $k_{max}\eta_{mag}$ & $k_{max}\eta_{kin}$ & $R_{-}$ & $R_L$ & $\Rl$ & $\vep_{mag}/\vep$ & $\vep_{kin}/\vep$ & $\mu=\nu $  & $k_{0}$ & \#  & $\Ceps$
  & $\sigma$ & $\rho_c(0)$ \\
  \hline
  NH1 & $256^3$ & 1.51 & 1.69 & 55.57 & 53.89 & 25.57 & 0.61 & 0.39& $0.004$  & 5 & 10 & 0.587 &0.005 & 0 \\
  NH2 & $256^3$ & 1.26 & 1.38 & 71.51 & 68.60 & 30.11 & 0.59& 0.41& $0.003$ & 5 & 10 & 0.530 &0.004 & 0\\
  NH3 & $512^3$ & 1.86 & 2.09 & 103.41 & 96.69 & 37.68 & 0.62& 0.38& $0.002$  & 5 & 10 & 0.468 &0.004 & 0\\
  NH4 & $512^3$ & 1.51 & 1.71 & 133.14 & 122.51 & 43.94 & 0.62& 0.38& $0.0015$  & 5 & 10 & 0.431 &0.004 & 0\\
  NH5 & $512^3$ & 1.29 & 1.47 & 161.35 & 151.76 & 50.73 & 0.63& 0.37& $0.0012$  & 5 & 10 & 0.394 &0.004 & 0\\
  NH6 & $1024^3$ & 2.28 & 2.55 & 192.40 & 168.28 & 54.44 & 0.61& 0.39& $0.001$ & 5 & 5 & 0.358 & 0.002 & 0\\
  NH7 & $1024^3$ & 1.76 & 1.98 & 259.58 & 232.10 & 65.42 & 0.62& 0.38& $0.0007$ & 5 & 5 & 0.358 & 0.002 & 0 \\
  NH8 & $1024^3$ & 1.40 & 1.56 & 354.30 & 301.71 & 76.73 & 0.61& 0.39& $0.0005$  & 5 & 5 & 0.323 & 0.002 & 0\\
  NH9 & $2048^3$ & 1.15 & 1.29 & 1071.44 & 823.58 & 134.73 & 0.61& 0.39& $0.00015$  & 5 & 1 & 0.279 & - & 0\\
  \hline
  CH06NH1 & $512^3$ & 2.02 & 2.23 & 94.39 & 113.02 & 49.29 & 0.60 & 0.40 & $0.002$ & 5 & 1 & 0.482 & - & 0.6 \\
  CH06NH2 & $512^3$ & 1.41 & 1.55 & 148.86 & 174.61 & 65.48 & 0.59 & 0.41& $0.0012$ & 5 & 5 & 0.417  & 0.003 & 0.6 \\
  CH06NH3 & $1024^3$ & 1.93 & 2.13  & 242.06 & 272.85 & 87.50 & 0.60 & 0.40& $0.0007$ & 5 & 1 & 0.365 & - & 0.6 \\
  CH06NH4 & $1024^3$ & 1.52 & 1.67 & 325.62 & 365.54 & 104.45 & 0.59 & 0.41& $0.0005$ & 5 & 1 & 0.341  & - & 0.6 \\
  CH06NH5 & $1024^3$ & 1.16 & 1.29 & 450.01 & 515.23 & 127.09 & 0.61& 0.39& $0.00035$ & 5 & 1 & 0.313 & - & 0.6 \\
  \end{tabular}
  \end{center}
 \caption{Specifications of decaying simulations \cite{data15a,Linkmann15a} for magnetic fields with negligible 
initial magnetic helicity.
 $R_L$ denotes the integral-scale Reynolds number,
$R_{\lambda}$ the Taylor-scale Reynolds number,
 $R_{-}$ the generalized Reynolds number, $\mu$ the
 magnetic resistivity, $k_0$ the peak wavenumber of the initial energy spectra,
 $k_{max}$ the largest resolved wavenumber, 
\blue{$\eta_{mag} = (\mu^3/\vep_{mag})^{1/4} $ and $\eta_{kin} = (\nu^3/\vep_{kin})^{1/4}$ the magnetic
and kinetic Kolmogorov microscales, respectively, at the peak of total dissipation},
 \# the ensemble size, $\Ceps$ the dimensionless total dissipation rate, $\sigma$ the
 standard error on $\Ceps$ and $\rho_c(0)$ the initial relative cross helicity. All Reynolds numbers are measured
 at the peak of total dissipation.
 }
 \label{tbl:simulations_nh}
 \end{table}

 \begin{table}[]
 \begin{center}
 \begin{tabular}{cccccccccccccc}
   Run id & $N^3$  & $k_{max}\eta_{mag}$ & $k_{max}\eta_{kin}$ & $R_{-}$ & $R_L$ & $\Rl$ & $\vep_{mag}/\vep$ & $\vep_{kin}/\vep$ & $\mu = \nu $ & $\Ceps$ 
  & $\sigma_{\Ceps}$ & $t/T$ \\
  \hline
ND1 & 128 & 3.22 & 2.53 & 27.93 & 78.68 & 42.65 & 0.28 & 0.72& 0.01 & 0.280 & 0.0028 & 30 \\ 
ND2 & 128 & 2.59 & 2.52 & 35.00 & 90.97 & 48.24 & 0.47  & 0.53& 0.009 & 0.317 & 0.0065 & 26 \\ 
ND3 & 128 & 2.48 & 2.25 & 38.31 & 106.34 & 54.22 & 0.40& 0.60 & 0.008 & 0.269 & 0.008 & 14 \\ 
ND4 & 128 & 2.12 & 2.12 & 45.63 & 119.92 & 59.59 & 0.50 & 0.50& 0.007 & 0.285 & 0.0021 & 27 \\ 
ND5 & 128 & 1.85 & 1.93 & 53.12 & 137.38 & 65.32 & 0.54 & 0.46& 0.006 & 0.290 & 0.0057 & 21 \\ 
ND6 & 128 & 1.46 & 1.58 & 66.21 & 173.33 & 75.83 & 0.58& 0.42 & 0.0045 & 0.285 & $10^{-5}$ & 10 \\ 
ND7 & 256 & 2.68 & 3.10 & 81.34 & 209.77 & 87.33 & 0.64& 0.36& 0.004 & 0.283 & 0.0029 & 10 \\ 
ND8 & 256 & 2.14 & 2.53 & 114.94 & 284.48 & 105.29 & 0.66 & 0.34& 0.003 & 0.272 & 0.015 & 17 \\ 
ND9 & 256 & 1.74 & 2.05 & 123.08 & 330.73 & 115.83 & 0.66 & 0.34& 0.0023 & 0.260 & 0.0072 & 18 \\ 
ND10 & 256 & 1.44 & 1.75 & 169.47 & 447.34 & 137.50 & 0.69 & 0.31& 0.0018 & 0.255 & 0.0025 & 27 \\ 
ND11 & 512 & 1.84 & 2.31 & 301.07 & 834.51 & 196.91 & 0.71 & 0.29& 0.001 & 0.239 & 0.015 & 16 \\ 
ND12 & 512 & 1.56 & 1.96 & 345.32 & 968.76 & 211.51 & 0.71& 0.29& 0.0008 & 0.238 & 0.0017 & 17 \\ 
ND13 & 512 & 1.45 & 1.82 & 359.68 & 1017.72 & 219.99 & 0.71&0.29 & 0.00073 & 0.235 & 0.0025 & 12 \\ 
ND14 & 528 & 1.37 & 1.72 & 454.79 & 1218.98 & 234.97 & 0.71& 0.29& 0.00067 & 0.231 & 0.0072 & 10 \\ 
ND15 & 1024 & 2.18 & 2.74 & 629.44 & 1593.58 & 236.97 & 0.71& 0.29& 0.0005 & 0.230 & 0.0041 & 12 \\ 
ND16 & 1024 & 1.51 & 1.86 & 919.16 & 2538.24 & 293.12  & 0.70 & 0.30 & 0.0003 & 0.226 & 0.0043 & 9 \\ 
  \hline
  SF1&$256$ & 1.71 &  2.04 & 152.95 & 428.57 & 140.64 & 0.67 &0.33 & 0.0035 & 0.240 & 0.0056 & 40\\ 
  SF2&$256$ & 1.52 &  1.85 &197.87  &481.87 & 144.61 & 0.69 & 0.31& 0.003 & 0.251 & 0.01 & 40 \\ 
  SF3&$256$ & 1.23 &  1.51& 239.91 &620.40 & 165.39 & 0.69 & 0.31& 0.0023 & 0.245 & 0.0028 & 40 \\ 
  SF4&$256$ & 1.01 & 1.25 & 315.74 & 812.38& 184.64 & 0.70& 0.30& 0.0018 &0.246 &0.0022 & 40 \\ 
  SF5&$512$ &1.69 & 2.11 & 392.42 & 1039.16& 213.55 & 0.71 & 0.29& 0.0014 & 0.238 & 0.0035 & 30 \\ 
  SF6&$512$ & 1.31 & 1.65 & 528.32 &1443.83& 251.76 & 0.72 & 0.28& 0.001 & 0.230 & 0.0006 & 30 \\ 
  \hline
HF1 & 256 & 1.52 & 1.75 & 135.81 & 385.19 & 123.22 & 0.64& 0.36& 0.0018 & 0.260 & 0.0055 & 15 \\ 
HF2 & 256 & 1.04 & 1.29 & 262.66 & 735.98 & 184.03 & 0.70& 0.30& 0.0014 & 0.237 & 0.0028 & 23 \\ 
HF3 & 512 & 0.96 & 1.17 & 613.83 &1718.40 & 267.67 & 0.69& 0.31& 0.0006  & 0.228 & 0.0075 & 15 \\ 
  \hline
  \end{tabular}
  \end{center}
 \caption{Specifications of stationary simulations. 
ND, SF and HF refer to the forcing schemes $\vec{f}_1$, $\vec{f}_2$ and $\vec{f}_3$, respectively. 
 $R_L$ denotes the integral-scale Reynolds number,
 $R_{\lambda}$ the Taylor-scale Reynolds number,
 $R_{-}$ the generalized Reynolds number given in \eqref{eq:gen_Rey}, $\mu$ the
 magnetic resistivity, 
 $k_{max}$ the largest resolved wavenumber, 
\blue{$\eta_{mag} = (\mu^3/\vep_{mag})^{1/4} $ and $\eta_{kin} = (\nu^3/\vep_{kin})^{1/4}$ the magnetic
and kinetic Kolmogorov microscales, respectively},
 $\Ceps$ the dimensionless total dissipation rate defined
 in \eqref{eq:ceps_defn}, $\sigma_{\Ceps}$ the
 standard error on $\Ceps$ and $t/T$ the run time in stationary state in
units of large-eddy turnover time $T=L_u/U$. All Reynolds numbers are time averages.
 }
 \label{tbl:simulations_forced}
 \end{table}

\subsection{Decaying MHD turbulence}
\label{sec:decay}
In this section we review the numerical results of Ref.~\cite{Linkmann15a}.
In order to compare data for nonstationary systems at different generalized Reynolds numbers,
we measure all quantities at the peak of dissipation \cite{Mininni09,Dallas13a}. Ensembles of up to 
$10$ runs per data point were used in order to calculate statistics. Evidently, a larger ensemble would 
be desirable, however, especially at high resolution the computational cost of a single run is already 
substantial. Therefore we compromised on the ensemble size in favor of running larger simulations, which is
essential for the present study. Four series of simulations were carried out which differ between 
each other in the initial values of the ideal invariants $H_m$ and $H_c$. Series H refers to 
a series with maximal initial $H_m$ while $H_c =0$, series CH06H was initialized with maximal $H_m$ and 
relative cross-helicity $\rho_c=H_c/(UB) =0.6$. Series NH and CH06NH label simulations initialized with $H_m=0$ and
differ in the initial level of $\rho_c$, with $\rho_c=0$ for series NH and $\rho_c=0.6$ for series CH06NH.   
All simulations resolve the magnetic and kinetic Kolmogorov scales
\blue{$\eta_{mag}=(\mu^3/\vep_{mag})^{1/4}$ and 
$\eta_{kin}= (\nu^3/\vep_{kin})^{1/4}$}, that is
$k_{max}\eta_{mag,kin} \geqslant 1$. 
Further details of series H and CH06H are given in Tbl.~\ref{tbl:simulations_h} while details corresponding to 
series NH and CH06NH are shown in Tbl.~\ref{tbl:simulations_nh}.  

\begin{figure}[!t]
 \begin{center}
\hspace{-2em} 
 \includegraphics[width=0.5\textwidth]{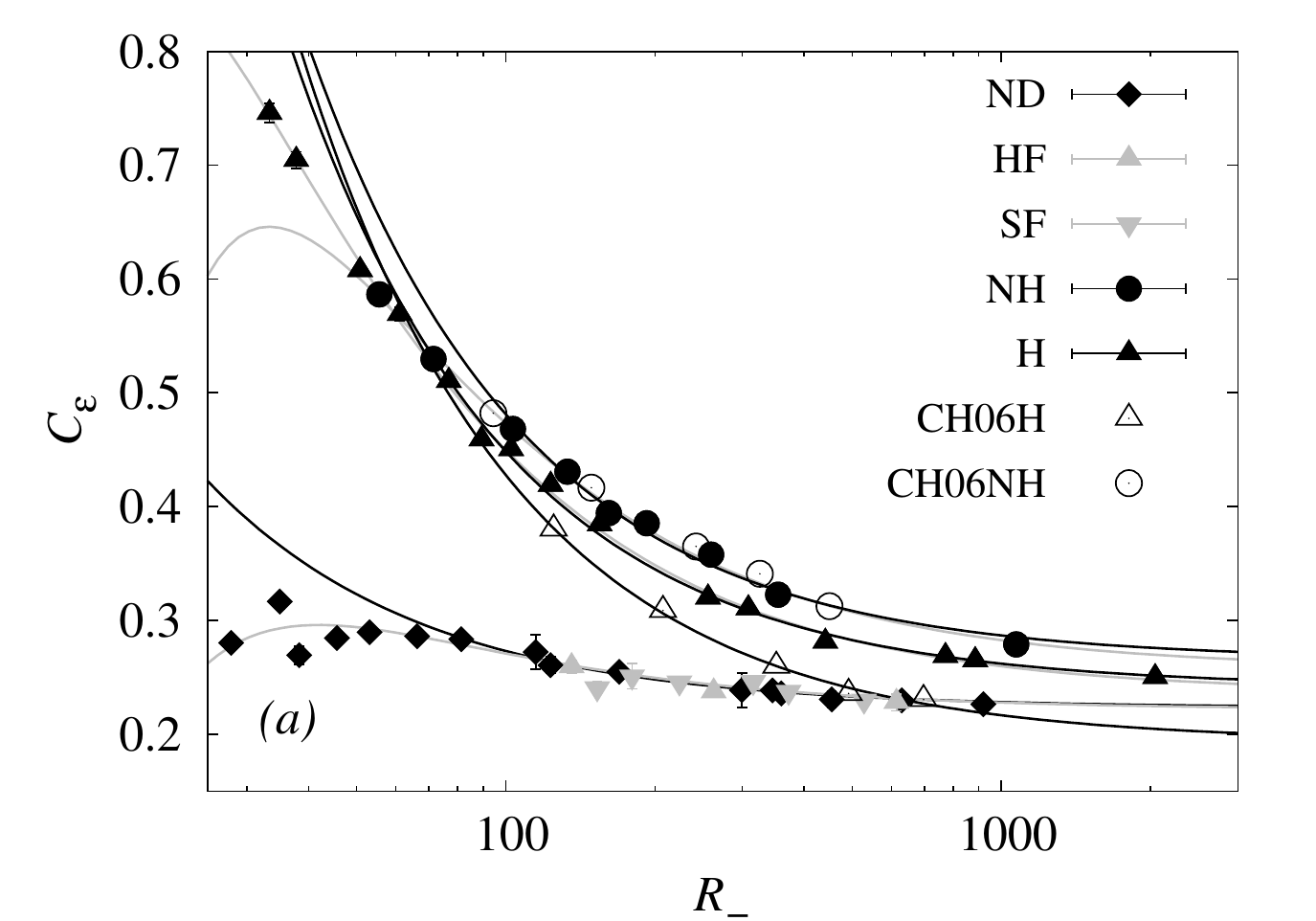}
\hspace{-2em} 
\includegraphics[width=0.5\textwidth]{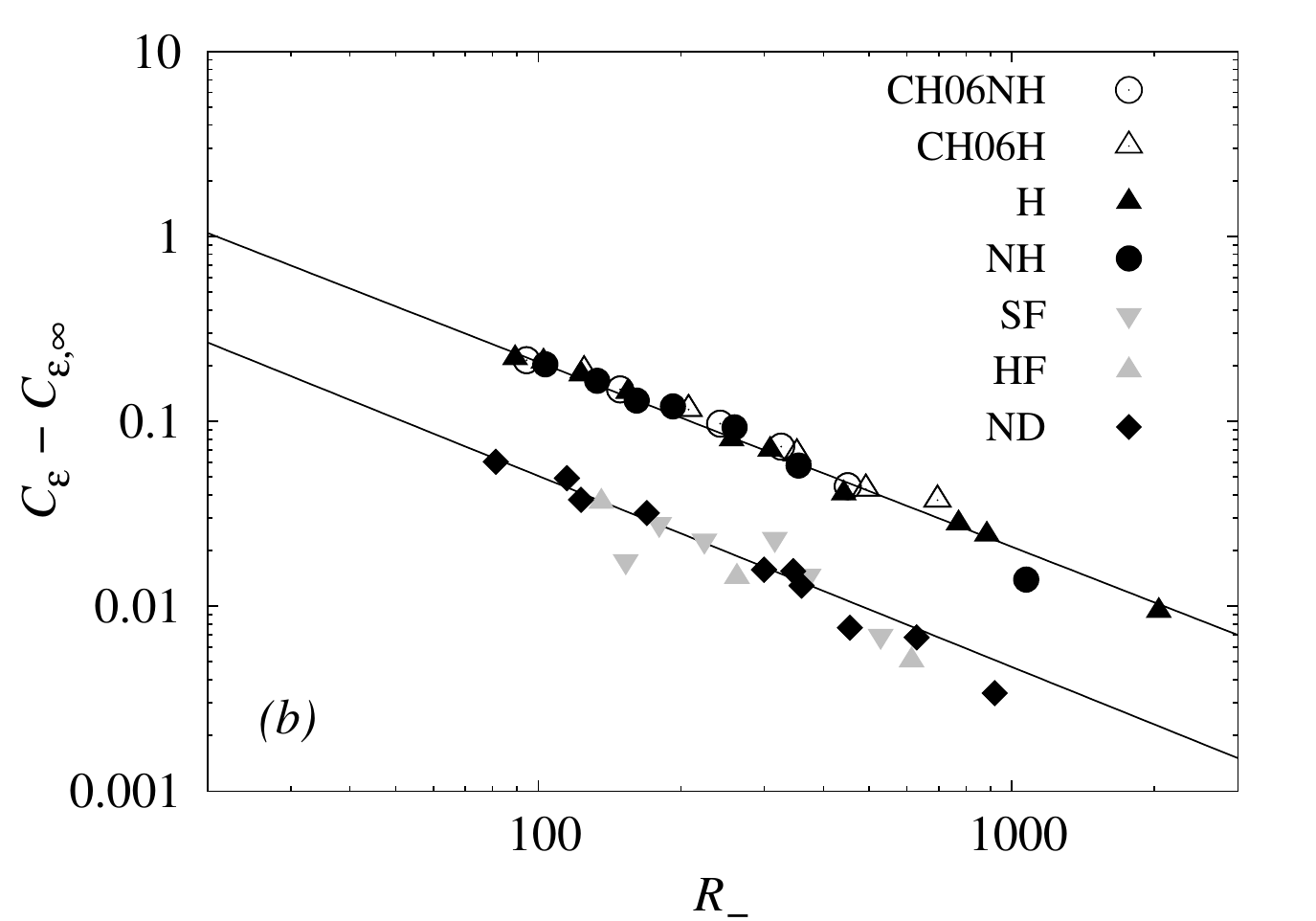}
 \caption{
  \blue{(a)} Eq.~\eqref{eq:model} fitted to the different datasets H, NH, CH06H, CH06NH (free decay) and ND (stationary). 
  Data from the stationary series SF and HF are shown for comparison purposes.
  The black lines refer to fits using Eq.~\eqref{eq:model} up to
  first order, while the gray lines use Eq.~\eqref{eq:model} up to second order in
  $1/R_-$. 
  The error bars show one standard error. 
  As can be seen, the respective asymptotes differ between the data sets H, NH, CH06H, CH06NH and ND, 
  while data from series SF and HF is compatible with data from series ND.
  \blue{(b) Fit of the expression $C/R_-^n$ corresponding to eq.~\eqref{eq:model} after 
   subtraction of the asymptote $\Cinf$ on a logarithmic scale to
   the datasets H (free decay, top line) and ND (stationary, bottom line). 
   Data from the series NH, CH06H, CH06NH, SF and HF are shown for comparison purposes. The 
   resulting exponents are $n=1.00 \pm 0.01$ for free decay and $n=1.00 \pm 0.02$ for the stationary 
   case.} 
 }
 \label{fig:comp_all}
 \end{center}
\end{figure}

Figure \ref{fig:comp_all}(a) shows fits of Eq.~\eqref{eq:model} to DNS
data for datasets that differ in the initial
value of $H_m$ and $H_c$. As can be seen, eq.~\eqref{eq:model} fits the data very well.
For the series H runs and for $R_- > 70$
it is sufficient to consider terms of first order in $R_-$,
while for the series NH the first-order approximation is valid for $R_- > 100$, 
\blue{as can be seen from Fig~\ref{fig:comp_all}(b), where 
a function of the form $C/R_-^n$ was fitted to data from series H
for $R_- > 70$ after subtraction of the asymptote $\Cinf$. 
The fit resulted in the value $n = 1.00 \pm 0.01$ for the exponent, 
and the data from series NH, CH06H and CH06NH are consistent with this 
result. Furthermore, Fig.~\ref{fig:comp_all}(a) shows that }
the cross-helical CH06H runs gave
consistently lower values of $\Ceps$ compared to the series H runs, while
little difference was observed between series CH06NH and NH.
The asymptotes were $\Cinf=0.241 \pm 0.008$
for the H series, $\Cinf=0.265 \pm 0.013$
for the NH series, $\Cinf=0.193 \pm 0.006$ for the CH06H series
and $\Cinf=0.268 \pm 0.005$ for the CH06NH series.

As predicted by the qualitative theoretical arguments outlined previously,
the measurements show that the asymptote calculated from the nonhelical runs
is larger than for the helical case,
as can be seen in Fig.~\ref{fig:comp_all}.
The asymptotes of the series H and NH do not lie within one standard
error of one another.
Simulations carried out with $H_c \neq 0$ suggest little difference
in $\Ceps$ for magnetic fields with initially zero magnetic helicity. For initially
helical magnetic fields $\Ceps$ is further quenched if $H_c \neq 0$.
In view of nonuniversality, an even larger variance
of $\Cinf$ can be expected once other parameters such as external forcing,
plasma $\beta$, $Pm$, etc., are taken into account. Here, attention is restricted
to nonuniversality related to different levels of cross- and magnetic helicity. 
The effect of external forcing will be analyzed in the next section. 

\subsection{Stationary MHD turbulence}
\label{sec:stationary}
In this case measurements are taken after the simulations
have reached a stationary state. The value of $\Ceps$ and the corresponding statistics 
for each data point are calculated from time-series obtained by evolving the 
stationary simulations for a minimum of $9$ large-eddy turnover times, as specified in 
Tbl.~\ref{tbl:simulations_forced}. 
All runs of the series ND satisfy $k_{max}\eta_{mag,kin} \geqslant 1.37$ 
and as such are sufficiently resolved.
The runs SF4, HF3 and HF4 are marginally resolved, and 
we point out that the fitting procedure only involved data obtained from the ND series. The 
data points obtained from the series SF and HF are included for comparison purposes. 
Further details of the stationary simulations are given in Tbl.~\ref{tbl:simulations_forced}.   
All helicities are initially negligible and remain so during the evolution of the simulations. 

Figure \ref{fig:comp_all} (a) shows error-weighted
fits of Eq.~\eqref{eq:model} to DNS data obtained from series ND. As can be seen,
Eq.~\eqref{eq:model} fits the data well, provided terms of 
second order in $R_{-}$ are included \blue{to accommodate the data points at
low $R_{-}$}. For $R_{-} > 80$, it is sufficient to consider 
terms of first order in $R_{-}$ only. \blue{The power law scaling to first order in 
$R_{-}$ is shown in further detail in Fig.~\ref{fig:comp_all} (b), 
where a function of the form $C/R_-^n$ was fitted to data from series ND 
for $R_- > 80$ after subtraction of the asymptote $\Cinf$. 
The fit resulted in the value $n = 1.00 \pm 0.02$ for the exponent, 
thus confirming that Eq.~\eqref{eq:model} describes the variation 
of $\Ceps$ at moderate to high $R_-$  well already at first order 
in $R_{-}$. This is similar to results in isotropic hydrodynamic
turbulence, where the corresponding hydrodynamic
equation ${\Ceps}_u = {\Cinf}_u + C_u/R_L$ agreed well with the 
data for $R_L > 80$ \cite{McComb15a}. In this context we point out that 
the lowest value of $R_L$ in Ref.~\cite{McComb15a} was $R_L = 81.5$.    
For $R_{-} \leqslant 80$ we find the data to be consistent 
with Eq.~\eqref{eq:model} once terms of second order in $R_{-}$
are taken into account. However, it can be very difficult to extract 
two power laws clearly from numerical data, especially if the leading and the 
subleading coefficients are of opposite sign \cite{Biferale04,Mitra05}. 
This is the case here, the subleading coefficient $D$ is always 
negative while the leading coefficient $C$ is always positive.    
}

Figure \ref{fig:comp_all} also shows that the result is independent of the 
forcing scheme, as the datasets obtained from simulations 
using the three different forcing functions are consistent with each other. 
This is likely to change if the strategy of energy input is fundamentally
changed, for example if the helical content and/or the characteristic scale 
of the force are altered, or if 
an electromagnetic force is used. We will come back to this point in Sec.~\ref{sec:conclusions}.
The independence of $\Ceps$ of the 
forcing scheme established here only shows independence of
the specific implementation of the forcing. 
The asymptote has been calculated to be $\Cinf=0.223 \pm 0.003$, 
where the error is obtained from the fit. 

A comparison of the measured values for $\Cinf$ 
obtained from the 
different simulation series is provided in Tbl.~\ref{tbl:cinf_summary}.     
Comparing the results from stationary and decaying simulations with 
the same level of helicities could shed some light on the effect of 
external forces in the context of nonuniversality. As such, we compare
the measured value of $\Cinf=0.223$ to the value calculated for the series 
of decaying simulations NH, which results in a difference of about $12 \%$.
Since $\Cinf$ does not depend explicitly on the external forces, the difference 
between the measured values may originate from dynamical effects. In relation to the 
effect of initial cross- and magntic helicities on the value of $\Cinf$ discussed in Sec.~\ref{sec:decay}, 
we expect further variance in the measured value of $\Cinf$ depending on the 
level of cross- and magnetic helicities of the external forces.
As can be seen from a comparison of the curves shown in Fig.~\ref{fig:comp_all}, 
the coefficient $C$ in Eq.~\eqref{eq:model} is {\em not} the same for the stationary and 
decaying cases. This is expected since $C^\pm$ and hence $C$ 
depend explicitly on the energy input, as can be seen from Eq.~\eqref{eq:c+}.

 \begin{table}[h]
 \begin{center}
 \begin{tabular}{lccl}
   Series id & $\Cinf$  & $\sigma_{\Cinf}$  &description \\
  \hline
   H & 0.241 & 0.008 &  decaying, helical \\
   NH & 0.265 & 0.013 &  decaying, nonhelical \\
   CH06H & 0.193 & 0.006 &  decaying, helical, cross-helical \\
   CH06NH & 0.268 & 0.005 &  decaying, nonhelical, cross-helical \\
   ND & 0.223 & 0.003 &  stationary, nonhelical \\
   Ref.~\cite{McComb15a} & 0.468 & 0.006 &  non-conducting, stationary 
  \end{tabular}
  \end{center}
 \caption{
Summary of measured values for the asymptotic dimensionless dissipation rate $\Cinf$  
from the different simulation series. For comparison the measured values
from a series of DNSs of stationary isotropic hydrodynamic turbulence \cite{McComb15a} are included.
The error on $\Cinf$  is denoted by $\sigma_{\Cinf}$ and refers to the error obtained
from the error-weighted fitting procedure.
 }
 \label{tbl:cinf_summary}
 \end{table}

\blue{Some further observations can be made from a comparison of the datasets
concerning the variance of the magnetic and kinetic contributions, $\vep_{mag}$ and
$\vep_{kin}$, to the total energy dissipation rate.  The fractions $\vep_{mag}/\vep$
and $\vep_{kin}/\vep$ are given columns 8 and 9, respectively, of
Tbls.~\ref{tbl:simulations_h}-\ref{tbl:simulations_forced} for the different
datasets. For the series H and CH06H it can be seen that the kinetic
dissipation fraction $\vep_{kin}/\vep$ grows with increasing $R_-$ (or $R_L$), of
course at the expense of the magnetic dissipation fraction $\vep_{mag}/\vep$.  The
reason for this behaviour could be connected with the inverse cascade of
magnetic helicity becoming more efficient at higher $R_-$ resulting in a higher
residual inverse transfer of magnetic energy and thus slightly less magnetic
energy to be dissipated at the small scales. This interpretation is supported
by the observation that no such variation is present for the nonhelical series
NH and CH06NH, where we measure $\vep_{mag}/\vep \simeq 0.6$ and $\vep_{mag}/\vep
\simeq 0.4$ for all data points.  For all four datasets corresponding to freely
decaying MHD turbulence the magnetic dissipation fraction is always higher than
the kinetic dissipation fraction, $\vep_{mag}/\vep > \vep_{kin}/\vep$. }

\blue{
The stationary datasets show yet
a different variation of $\vep_{mag}/\vep$ and $\vep_{kin}/\vep$ with $R_-$.  For $R_-
< 45$ we find $\vep_{mag}/\vep < \vep_{kin}/\vep$, while for $R_- > 46$ the results are
similar to free decay with $\vep_{mag}/\vep > \vep_{kin}/\vep$. The magnetic
dissipation fraction increases with $R_-$ from $\vep_{mag}/\vep \simeq 0.3$ to
$\vep_{mag}/\vep \simeq 0.7$, where it appears to reach a plateau. 
The fluctuating magnetic field is maintained by the
velocity field fluctuations through a nonlinear dynamo process in the present 
DNSs of stationary MHD turbulence. Hence in the statistically stationary state
$\vep_{mag}$ is in balance with the dynamo term 
$\langle \vec{b}\cdot (\vec{b}\cdot \nabla) \vec{u} \rangle$, and the measured
values of $\vep_{mag}/\vep$ and $\vep_{kin}/\vep$ imply that at lower Reynolds number 
the nonlinear dynamo is less efficient in maintaining small-scale magnetic field 
fluctuations than at higher Reynolds number. Similar conclusions have been reached 
in a study of the magnetic Prandtl number dependence of the ratio $\vep_{kin}/\vep_{mag}$ 
\cite{Brandenburg14b}, where the efficiency of the dynamo at different values of  
$Pm$ was linked to measured values of $\vep_{kin}/\vep_{mag}$.}

\section{Conclusions}
\label{sec:conclusions}
The behavior of the dimensionless dissipation coefficient $\Ceps$ in 
homogeneous MHD turbulence with $Pm=1$ and no background magnetic field 
is well described by 
\beq
\Ceps = \Cinf + \frac{C}{R_-} + \frac{D}{R_-^2} + O(R_-^{-3}) \ .
\label{eq:model2}
\eeq
This equation was derived from the energy balance equation for $\vec{z}^\pm$ in real space (the vKHE)
by outer asymptotic expansions in powers of $1/R_{\mp}$, {\em leading necessarily to a 
large-scale description of the behavior of the dimensionless dissipation rate}. 
The approximative equation \eqref{eq:model2} has been shown to agree well with data obtained from 
medium to high resolution DNSs of both 
decaying MHD turbulence at the peak of dissipation and statistically steady MHD turbulence 
sustained by large-scale forcing. 
The measurements for $\Cinf$ ranged from $0.193 \leqslant \Cinf \leqslant 0.268$
between the different series of simulations. 
Interestingly, the measured values of $\Cinf$ for MHD are
smaller than the measured value of $\Cinf \simeq 0.5$ in hydrodynamic
turbulence obtained both from numerical simulations and experiments
\cite{Sreenivasan84,Jimenez93,Wang96,Yeung97,Sreenivasan98,Cao99,Pearson02,Kaneda03,Pearson04a,Donzis05,Bos07,Yeung12,McComb15a,Yeung15}, 
suggesting less energy transfer across scales in MHD turbulence compared to hydrodynamics. 

The asymptote in the limit $R_{-} \to \infty$ originates from the sum of the nonlinear terms
in the momentum and induction equations, that is, it measures the total transfer flux, which 
is expected to depend on the values of the ideal invariants.
As predicted, the values of the respective asymptotes from the datasets differ,
suggesting a dependence of $\Cinf$ on different values of the helicities,
and thus a connection to questions of universality in MHD turbulence.
For maximally helical magnetic fields $\Cinf$ is smaller than for nonhelical fields. This is expected
from the inverse cascade of magnetic helicity. The dependence of $\Cinf$ on the remaining
ideal invariant, the cross-helicity, is more complex. Since $\Cinf$ describes the 
flux of total energy across the scales, this flux is expected to diminish for increasing 
cross-helicity. This is indeed the case for helical magnetic fields, where 
$\Cinf$ depends on the cross-helicity in the expected way. Surprisingly, for nonhelical 
magnetic fields $\Cinf$ does {\em not} depend on the cross-helicity. 
This is consistent with the asymmetric effect of 
the cross-helicity on forward and inverse fluxes of total energy  
suggested by the analysis of triad interactions in Ref.~\cite{Linkmann16a}, 
where high levels of cross-helicity were found to quench 
forward transfer more than inverse transfer. A similar effect can be inferred from 
predictions obtained from statistical mechanics \cite{Frisch75}, where the simultaneous 
presence of cross- and magnetic helicities resulted in inverse transfers of {\em both}
magnetic and kinetic energy. In this case, the forward flux of total energy should be 
lower than for all other cases, which is consistent with the numerical results presented here.     
Concerning stationary nonhelical MHD turbulence, we found that $\Cinf$ differed by about $12 \%$ 
from the value measured for nonhelical decaying turbulence (series NH) at the peak of dissipation. 
According to the results from the asymptotic analysis, there is no explicit dependence of $\Cinf$ on 
the external force. As such, the difference in the measured value of $\Cinf$ between the stationary and the 
decaying systems may be due to dynamical effects, which may be interpreted as further support for 
nonuniversal values of $\Ceps$. The approach to the asymptote is predicted to differ between 
decaying and stationary systems due to the explicit dependence of the coefficient $C$ in 
Eq.~\eqref{eq:model} on the forcing. This is indeed observed in the simulations. 
 
The numerical results showed that $\Cinf$ is universal with respect to different 
forcing schemes applied to the same field in the 
same wavenumber range, thus confirming that the particular 
functional form and stochasticity of a large-scale force is 
irrelevant to the small-scale turbulent dynamics as long as the ideal invariants remain the same 
for the different forcing schemes. However, as mentioned in Sec.~\ref{sec:numerics}, this is 
expected to change if the strategy of energy input is changed. The effect of large-scale 
magnetic forcing on the scaling of $\vep$ with different rms 
quantities was investigated recently \cite{Alexakis13}. Numerical results showed 
that $\vep \sim U^3/L_{\vec{f}_u}$ for a large region of parameter space even in the presence of 
electromagnetic forces. Only once the large-scale magnetic field became very strong
a different scaling related to the dominance of magnetic shear over mechanical shear was found:
$\vep \sim BU^2/L_{\vec{f}_b}$.   
Differences may also be expected for forces applied at smaller scales.
      The analysis presented here relies on taking outer asymptotic expansions
      of all scale-dependent functions in the vKHE, including 
      the energy input from the forcing. Here it was crucial to assume that the system 
      was forced at the large scales, as the limit of infinite Reynolds number was defined
      as energy input at the lowest wavenumbers $k \to 0$ and removal of energy at the largest 
      wavenumbers $k \to \infty$. This clearly precludes the application of the present analysis
      to situations where the system is forced at intermediate or small scales. 
      Therefore, it can be expected that systems forced at intermediate scales deviate 
      from the $1/R_{-}$-scaling of $\Ceps$. 
For hydrodynamics, this is the case \cite{Doering09,Biferale04,Mazzi04}. 
Furthermore, experimental and numerical results for nonstationary flows 
\cite{Valente14,Vassilicos15} 
suggest even further variance possibly due to the influence of the 
time-derivative of the second-order structure function in the vKHE. 
 
The results presented here were restricted to homogeneous 
MHD turbulence at $Pm =1$ without a mean magnetic field. 
In general, further variance in the measured value for $\Cinf$ is expected depending 
on e.g.~the magnetic Prandtl number \cite{Brandenburg14b} or the 
influence of a background magnetic field.
The presence of a background magnetic field, which leads to spectral anisotropy and the 
breakdown of the conservation of magnetic helicity \cite{Matthaeus82a}, will
introduce several difficulties to be overcome when generalizing the analytical approach.
The spectral flux will then depend on the direction of the mean
field \cite{Wan09,Wan12} and a more generalized description
and role for the magnetic helicity would be needed \cite{Berger97,Berger99}. 
Other questions concern the generalization of this approach to 
MHD flows with magnetic Prandtl numbers $Pm \neq 1$, the effect of compressive fluctuations 
or the influence of other vector field correlations on the dissipation rate and/or 
the approach to the asymptote as observed by Dallas and Alexakis \cite{Dallas13a}. 

\section{Acknowledgments}
We thank V. Dallas for useful discussions. 
A.~B. acknowledges support from the UK Science and Technology Facilities Council,
M.~L. received support from the UK Engineering and Physical Sciences
Research Council (EP/K503034/1) and
E.~E.~G. was supported by a University of Edinburgh Physics and Astronomy Fellowship.
This work has made use of the resources provided by the UK
National Supercomputing facility ARCHER, ({\tt http://www.archer.ac.uk}),
made available through the Edinburgh Compute and Data Facility (ECDF, \\
{\tt http://www.ecdf.ed.ac.uk}). The research leading to these results 
has received funding from the European
Union's Seventh Framework Programme (FP7/2007-2013) under grant agreement No.
339032.

\appendix
\section{Gauge independence of Equation \eqref{eq:els_hmag}} 
\label{app:els_hmag}
In order to prove that Eq.~\eqref{eq:els_hmag} is correct 
for an arbitrary choice of gauge, we first express the current
density $\vec{j}$ in terms of the vector potential $\vec{a}$ 
\beq
\vec{j} = \nabla \times (\nabla \times \vec{a}) 
= - \Delta \vec{a} + \nabla (\nabla \cdot \vec{a}) \ ,
\eeq 
which holds in any gauge. In Fourier space this relation becomes
\beq
\fvec{j}= k^2 \fvec{a} + i\vec{k}(i\vec{k}\cdot \fvec{a}), \
\eeq
hence one obtains
\beq
\int_\Omega d \vec{k} \ \left \langle \frac{1}{k^2} \fvec{j} \cdot \fvec{b}^* \right \rangle 
 = \int_\Omega d \vec{k}  \ 
  \left \langle \left(\fvec{a} + \frac{i\vec{k}}{k^2}(i\vec{k}\cdot \fvec{a})\right) 
                      \cdot \fvec{b}^* \right \rangle 
 =\int_\Omega d \vec{k}  \ 
  \left \langle \fvec{a} \cdot \fvec{b}^* \right \rangle = H_m \ ,
\eeq
since $\vec{b}$ is a solenoidal vector field. Equation \eqref{eq:els_hmag}
now follows by writing the Fourier coefficients of the magnetic field and the current density in the Els\"{a}sser
formulation
\beq
\fvec{b} = \frac{\fvec{z}^+-\fvec{z}^-}{2} 
\qquad \mbox{and} \qquad 
\fvec{j} = i\vec{k} \times \fvec{b} = i\vec{k} \times \frac{\fvec{z}^+-\fvec{z}^-}{2}   \ . 
\eeq

\bibliography{refs,refs1,refs2,wdm}

\begin{thebibliography}{74}%
\makeatletter
\providecommand \@ifxundefined [1]{%
 \@ifx{#1\undefined}
}%
\providecommand \@ifnum [1]{%
 \ifnum #1\expandafter \@firstoftwo
 \else \expandafter \@secondoftwo
 \fi
}%
\providecommand \@ifx [1]{%
 \ifx #1\expandafter \@firstoftwo
 \else \expandafter \@secondoftwo
 \fi
}%
\providecommand \natexlab [1]{#1}%
\providecommand \enquote  [1]{``#1''}%
\providecommand \bibnamefont  [1]{#1}%
\providecommand \bibfnamefont [1]{#1}%
\providecommand \citenamefont [1]{#1}%
\providecommand \href@noop [0]{\@secondoftwo}%
\providecommand \href [0]{\begingroup \@sanitize@url \@href}%
\providecommand \@href[1]{\@@startlink{#1}\@@href}%
\providecommand \@@href[1]{\endgroup#1\@@endlink}%
\providecommand \@sanitize@url [0]{\catcode `\\12\catcode `\$12\catcode
  `\&12\catcode `\#12\catcode `\^12\catcode `\_12\catcode `\%12\relax}%
\providecommand \@@startlink[1]{}%
\providecommand \@@endlink[0]{}%
\providecommand \url  [0]{\begingroup\@sanitize@url \@url }%
\providecommand \@url [1]{\endgroup\@href {#1}{\urlprefix }}%
\providecommand \urlprefix  [0]{URL }%
\providecommand \Eprint [0]{\href }%
\providecommand \doibase [0]{http://dx.doi.org/}%
\providecommand \selectlanguage [0]{\@gobble}%
\providecommand \bibinfo  [0]{\@secondoftwo}%
\providecommand \bibfield  [0]{\@secondoftwo}%
\providecommand \translation [1]{[#1]}%
\providecommand \BibitemOpen [0]{}%
\providecommand \bibitemStop [0]{}%
\providecommand \bibitemNoStop [0]{.\EOS\space}%
\providecommand \EOS [0]{\spacefactor3000\relax}%
\providecommand \BibitemShut  [1]{\csname bibitem#1\endcsname}%
\let\auto@bib@innerbib\@empty
\bibitem [{\citenamefont {Kolmogorov}(1941)}]{Kolmogorov41a}%
  \BibitemOpen
  \bibfield  {author} {\bibinfo {author} {\bibfnamefont {A.~N.}\ \bibnamefont
  {Kolmogorov}},\ }\href@noop {} {\bibfield  {journal} {\bibinfo  {journal} {C.
  R. Acad. Sci. URSS}\ }\textbf {\bibinfo {volume} {30}},\ \bibinfo {pages}
  {301} (\bibinfo {year} {1941})}\BibitemShut {NoStop}%
\bibitem [{\citenamefont {Iroshnikov}(1964)}]{Iroshnikov64}%
  \BibitemOpen
  \bibfield  {author} {\bibinfo {author} {\bibfnamefont {P.~S.}\ \bibnamefont
  {Iroshnikov}},\ }\href@noop {} {\bibfield  {journal} {\bibinfo  {journal}
  {Soviet Astronomy}\ }\textbf {\bibinfo {volume} {7}},\ \bibinfo {pages} {566}
  (\bibinfo {year} {1964})}\BibitemShut {NoStop}%
\bibitem [{\citenamefont {Kraichnan}(1965)}]{Kraichnan65a}%
  \BibitemOpen
  \bibfield  {author} {\bibinfo {author} {\bibfnamefont {R.~H.}\ \bibnamefont
  {Kraichnan}},\ }\href@noop {} {\bibfield  {journal} {\bibinfo  {journal}
  {Phys. Fluids}\ }\textbf {\bibinfo {volume} {8}},\ \bibinfo {pages} {1385}
  (\bibinfo {year} {1965})}\BibitemShut {NoStop}%
\bibitem [{\citenamefont {Goldreich}\ and\ \citenamefont
  {Sridhar}(1995)}]{Goldreich95}%
  \BibitemOpen
  \bibfield  {author} {\bibinfo {author} {\bibfnamefont {P.}~\bibnamefont
  {Goldreich}}\ and\ \bibinfo {author} {\bibfnamefont {S.}~\bibnamefont
  {Sridhar}},\ }\href@noop {} {\bibfield  {journal} {\bibinfo  {journal}
  {Astrophys. J.}\ }\textbf {\bibinfo {volume} {438}},\ \bibinfo {pages} {763}
  (\bibinfo {year} {1995})}\BibitemShut {NoStop}%
\bibitem [{\citenamefont {Boldyrev}(2005)}]{Boldyrev05a}%
  \BibitemOpen
  \bibfield  {author} {\bibinfo {author} {\bibfnamefont {S.}~\bibnamefont
  {Boldyrev}},\ }\href@noop {} {\bibfield  {journal} {\bibinfo  {journal}
  {ApJ}\ }\textbf {\bibinfo {volume} {626}},\ \bibinfo {pages} {L37} (\bibinfo
  {year} {2005})}\BibitemShut {NoStop}%
\bibitem [{\citenamefont {Boldyrev}(2006)}]{Boldyrev06}%
  \BibitemOpen
  \bibfield  {author} {\bibinfo {author} {\bibfnamefont {S.}~\bibnamefont
  {Boldyrev}},\ }\href@noop {} {\bibfield  {journal} {\bibinfo  {journal}
  {Phys. Rev. Lett.}\ }\textbf {\bibinfo {volume} {96}},\ \bibinfo {pages}
  {115002} (\bibinfo {year} {2006})}\BibitemShut {NoStop}%
\bibitem [{\citenamefont {Beresnyak}\ and\ \citenamefont
  {Lazarian}(2006)}]{Beresnyak06}%
  \BibitemOpen
  \bibfield  {author} {\bibinfo {author} {\bibfnamefont {A.}~\bibnamefont
  {Beresnyak}}\ and\ \bibinfo {author} {\bibfnamefont {A.}~\bibnamefont
  {Lazarian}},\ }\href@noop {} {\bibfield  {journal} {\bibinfo  {journal}
  {ApJL}\ }\textbf {\bibinfo {volume} {640}},\ \bibinfo {pages} {L175}
  (\bibinfo {year} {2006})}\BibitemShut {NoStop}%
\bibitem [{\citenamefont {Mason}\ \emph {et~al.}(2006)\citenamefont {Mason},
  \citenamefont {Cattaneo},\ and\ \citenamefont {Boldyrev}}]{Mason06}%
  \BibitemOpen
  \bibfield  {author} {\bibinfo {author} {\bibfnamefont {J.}~\bibnamefont
  {Mason}}, \bibinfo {author} {\bibfnamefont {F.}~\bibnamefont {Cattaneo}}, \
  and\ \bibinfo {author} {\bibfnamefont {S.}~\bibnamefont {Boldyrev}},\
  }\href@noop {} {\bibfield  {journal} {\bibinfo  {journal} {Phys. Rev. Lett.}\
  }\textbf {\bibinfo {volume} {97}},\ \bibinfo {pages} {255002} (\bibinfo
  {year} {2006})}\BibitemShut {NoStop}%
\bibitem [{\citenamefont {Gogoberidze}(2007)}]{Gogoberidze07}%
  \BibitemOpen
  \bibfield  {author} {\bibinfo {author} {\bibfnamefont {G.}~\bibnamefont
  {Gogoberidze}},\ }\href@noop {} {\bibfield  {journal} {\bibinfo  {journal}
  {Phys. Plasmas}\ }\textbf {\bibinfo {volume} {14}},\ \bibinfo {pages}
  {022304} (\bibinfo {year} {2007})}\BibitemShut {NoStop}%
\bibitem [{\citenamefont {Dallas}\ and\ \citenamefont
  {Alexakis}(2013{\natexlab{a}})}]{Dallas13a}%
  \BibitemOpen
  \bibfield  {author} {\bibinfo {author} {\bibfnamefont {V.}~\bibnamefont
  {Dallas}}\ and\ \bibinfo {author} {\bibfnamefont {A.}~\bibnamefont
  {Alexakis}},\ }\href@noop {} {\bibfield  {journal} {\bibinfo  {journal}
  {Phys. Fluids}\ }\textbf {\bibinfo {volume} {25}},\ \bibinfo {pages} {105106}
  (\bibinfo {year} {2013}{\natexlab{a}})}\BibitemShut {NoStop}%
\bibitem [{\citenamefont {Dallas}\ and\ \citenamefont
  {Alexakis}(2013{\natexlab{b}})}]{Dallas13b}%
  \BibitemOpen
  \bibfield  {author} {\bibinfo {author} {\bibfnamefont {V.}~\bibnamefont
  {Dallas}}\ and\ \bibinfo {author} {\bibfnamefont {A.}~\bibnamefont
  {Alexakis}},\ }\href@noop {} {\bibfield  {journal} {\bibinfo  {journal}
  {Phys. Rev. E}\ }\textbf {\bibinfo {volume} {88}},\ \bibinfo {pages} {063017}
  (\bibinfo {year} {2013}{\natexlab{b}})}\BibitemShut {NoStop}%
\bibitem [{\citenamefont {Wan}\ \emph {et~al.}(2012)\citenamefont {Wan},
  \citenamefont {Oughton}, \citenamefont {Servidio},\ and\ \citenamefont
  {Matthaeus}}]{Wan12}%
  \BibitemOpen
  \bibfield  {author} {\bibinfo {author} {\bibfnamefont {M.}~\bibnamefont
  {Wan}}, \bibinfo {author} {\bibfnamefont {S.}~\bibnamefont {Oughton}},
  \bibinfo {author} {\bibfnamefont {S.}~\bibnamefont {Servidio}}, \ and\
  \bibinfo {author} {\bibfnamefont {W.~H.}\ \bibnamefont {Matthaeus}},\
  }\href@noop {} {\bibfield  {journal} {\bibinfo  {journal} {J. Fluid Mech.}\
  }\textbf {\bibinfo {volume} {697}},\ \bibinfo {pages} {296} (\bibinfo {year}
  {2012})}\BibitemShut {NoStop}%
\bibitem [{\citenamefont {Schekochihin}\ \emph {et~al.}(2008)\citenamefont
  {Schekochihin}, \citenamefont {Cowley},\ and\ \citenamefont
  {Yousef}}]{Schekochihin08}%
  \BibitemOpen
  \bibfield  {author} {\bibinfo {author} {\bibfnamefont {A.~A.}\ \bibnamefont
  {Schekochihin}}, \bibinfo {author} {\bibfnamefont {S.~C.}\ \bibnamefont
  {Cowley}}, \ and\ \bibinfo {author} {\bibfnamefont {T.~A.}\ \bibnamefont
  {Yousef}},\ }in\ \href@noop {} {\emph {\bibinfo {booktitle} {{IUTAM
  {S}ymposium on {C}omputational {P}hysics and {N}ew {P}erspectives in
  {T}urbulence}}}},\ \bibinfo {editor} {edited by\ \bibinfo {editor}
  {\bibfnamefont {Y.}~\bibnamefont {Kaneda}}}\ (\bibinfo  {publisher}
  {Springer},\ \bibinfo {address} {Berlin},\ \bibinfo {year} {2008})\ pp.\
  \bibinfo {pages} {347--354}\BibitemShut {NoStop}%
\bibitem [{\citenamefont {Mininni}(2011)}]{Mininni11}%
  \BibitemOpen
  \bibfield  {author} {\bibinfo {author} {\bibfnamefont {P.~D.}\ \bibnamefont
  {Mininni}},\ }\href@noop {} {\bibfield  {journal} {\bibinfo  {journal} {Annu.
  Rev. Fluid Mech.}\ }\textbf {\bibinfo {volume} {43}},\ \bibinfo {pages} {377}
  (\bibinfo {year} {2011})}\BibitemShut {NoStop}%
\bibitem [{\citenamefont {Grappin}\ \emph {et~al.}(1983)\citenamefont
  {Grappin}, \citenamefont {Pouquet},\ and\ \citenamefont
  {L\'{e}orat}}]{Grappin83}%
  \BibitemOpen
  \bibfield  {author} {\bibinfo {author} {\bibfnamefont {R.}~\bibnamefont
  {Grappin}}, \bibinfo {author} {\bibfnamefont {A.}~\bibnamefont {Pouquet}}, \
  and\ \bibinfo {author} {\bibfnamefont {J.}~\bibnamefont {L\'{e}orat}},\
  }\href@noop {} {\bibfield  {journal} {\bibinfo  {journal} {Astron.
  Astrophys.}\ }\textbf {\bibinfo {volume} {126}},\ \bibinfo {pages} {51}
  (\bibinfo {year} {1983})}\BibitemShut {NoStop}%
\bibitem [{\citenamefont {{A. Pouquet and P. Mininni and D. Montgomery and A.
  Alexakis}}(2008)}]{Pouquet08}%
  \BibitemOpen
  \bibfield  {author} {\bibinfo {author} {\bibnamefont {{A. Pouquet and P.
  Mininni and D. Montgomery and A. Alexakis}}},\ }in\ \href@noop {} {\emph
  {\bibinfo {booktitle} {{IUTAM {S}ymposium on {C}omputational {P}hysics and
  {N}ew {P}erspectives in {T}urbulence}}}},\ \bibinfo {editor} {edited by\
  \bibinfo {editor} {\bibfnamefont {Y.}~\bibnamefont {Kaneda}}}\ (\bibinfo
  {publisher} {Springer},\ \bibinfo {year} {2008})\ pp.\ \bibinfo {pages}
  {305--312}\BibitemShut {NoStop}%
\bibitem [{\citenamefont {Beresnyak}(2011)}]{Beresnyak11}%
  \BibitemOpen
  \bibfield  {author} {\bibinfo {author} {\bibfnamefont {A.}~\bibnamefont
  {Beresnyak}},\ }\href@noop {} {\bibfield  {journal} {\bibinfo  {journal}
  {Phys. Rev. Lett.}\ }\textbf {\bibinfo {volume} {106}},\ \bibinfo {pages}
  {075001} (\bibinfo {year} {2011})}\BibitemShut {NoStop}%
\bibitem [{\citenamefont {Boldyrev}\ \emph {et~al.}(2011)\citenamefont
  {Boldyrev}, \citenamefont {Perez}, \citenamefont {Borovsky},\ and\
  \citenamefont {Podesta}}]{Boldyrev11}%
  \BibitemOpen
  \bibfield  {author} {\bibinfo {author} {\bibfnamefont {S.}~\bibnamefont
  {Boldyrev}}, \bibinfo {author} {\bibfnamefont {J.~C.}\ \bibnamefont {Perez}},
  \bibinfo {author} {\bibfnamefont {J.~E.}\ \bibnamefont {Borovsky}}, \ and\
  \bibinfo {author} {\bibfnamefont {J.~J.}\ \bibnamefont {Podesta}},\
  }\href@noop {} {\bibfield  {journal} {\bibinfo  {journal} {Astrophys. J.}\
  }\textbf {\bibinfo {volume} {741}},\ \bibinfo {pages} {L19} (\bibinfo {year}
  {2011})}\BibitemShut {NoStop}%
\bibitem [{\citenamefont {Grappin}\ and\ \citenamefont
  {M\"uller}(2010)}]{Grappin10}%
  \BibitemOpen
  \bibfield  {author} {\bibinfo {author} {\bibfnamefont {R.}~\bibnamefont
  {Grappin}}\ and\ \bibinfo {author} {\bibfnamefont {W.-C.}\ \bibnamefont
  {M\"uller}},\ }\href@noop {} {\bibfield  {journal} {\bibinfo  {journal}
  {Phys. Rev. E}\ }\textbf {\bibinfo {volume} {82}},\ \bibinfo {pages} {026406}
  (\bibinfo {year} {2010})}\BibitemShut {NoStop}%
\bibitem [{\citenamefont {Lee}\ \emph {et~al.}(2010)\citenamefont {Lee},
  \citenamefont {Brachet}, \citenamefont {Pouquet}, \citenamefont {Mininni},\
  and\ \citenamefont {Rosenberg}}]{Lee10}%
  \BibitemOpen
  \bibfield  {author} {\bibinfo {author} {\bibfnamefont {E.}~\bibnamefont
  {Lee}}, \bibinfo {author} {\bibfnamefont {M.~E.}\ \bibnamefont {Brachet}},
  \bibinfo {author} {\bibfnamefont {A.}~\bibnamefont {Pouquet}}, \bibinfo
  {author} {\bibfnamefont {P.~D.}\ \bibnamefont {Mininni}}, \ and\ \bibinfo
  {author} {\bibfnamefont {D.}~\bibnamefont {Rosenberg}},\ }\href@noop {}
  {\bibfield  {journal} {\bibinfo  {journal} {Phys. Rev. E}\ }\textbf {\bibinfo
  {volume} {81}},\ \bibinfo {pages} {016318} (\bibinfo {year}
  {2010})}\BibitemShut {NoStop}%
\bibitem [{\citenamefont {Servidio}\ \emph {et~al.}(2008)\citenamefont
  {Servidio}, \citenamefont {Matthaeus},\ and\ \citenamefont
  {Dmitruk}}]{Servidio08}%
  \BibitemOpen
  \bibfield  {author} {\bibinfo {author} {\bibfnamefont {S.}~\bibnamefont
  {Servidio}}, \bibinfo {author} {\bibfnamefont {W.~H.}\ \bibnamefont
  {Matthaeus}}, \ and\ \bibinfo {author} {\bibfnamefont {P.}~\bibnamefont
  {Dmitruk}},\ }\href@noop {} {\bibfield  {journal} {\bibinfo  {journal} {Phys.
  Rev. Lett.}\ }\textbf {\bibinfo {volume} {100}},\ \bibinfo {pages} {095005}
  (\bibinfo {year} {2008})}\BibitemShut {NoStop}%
\bibitem [{\citenamefont {Frisch}\ \emph {et~al.}(1975)\citenamefont {Frisch},
  \citenamefont {Pouquet}, \citenamefont {L\'{e}orat},\ and\ \citenamefont
  {Mazure}}]{Frisch75}%
  \BibitemOpen
  \bibfield  {author} {\bibinfo {author} {\bibfnamefont {U.}~\bibnamefont
  {Frisch}}, \bibinfo {author} {\bibfnamefont {A.}~\bibnamefont {Pouquet}},
  \bibinfo {author} {\bibfnamefont {J.}~\bibnamefont {L\'{e}orat}}, \ and\
  \bibinfo {author} {\bibfnamefont {A.}~\bibnamefont {Mazure}},\ }\href@noop {}
  {\bibfield  {journal} {\bibinfo  {journal} {J. Fluid Mech.}\ }\textbf
  {\bibinfo {volume} {68}},\ \bibinfo {pages} {769} (\bibinfo {year}
  {1975})}\BibitemShut {NoStop}%
\bibitem [{\citenamefont {Pouquet}\ \emph {et~al.}(1976)\citenamefont
  {Pouquet}, \citenamefont {Frisch},\ and\ \citenamefont
  {L{\'e}orat}}]{Pouquet76}%
  \BibitemOpen
  \bibfield  {author} {\bibinfo {author} {\bibfnamefont {A.}~\bibnamefont
  {Pouquet}}, \bibinfo {author} {\bibfnamefont {U.}~\bibnamefont {Frisch}}, \
  and\ \bibinfo {author} {\bibfnamefont {J.}~\bibnamefont {L{\'e}orat}},\
  }\href@noop {} {\bibfield  {journal} {\bibinfo  {journal} {J. Fluid Mech.}\
  }\textbf {\bibinfo {volume} {77}},\ \bibinfo {pages} {321} (\bibinfo {year}
  {1976})}\BibitemShut {NoStop}%
\bibitem [{\citenamefont {Pouquet}\ and\ \citenamefont
  {Patterson}(1978)}]{Pouquet78}%
  \BibitemOpen
  \bibfield  {author} {\bibinfo {author} {\bibfnamefont {A.}~\bibnamefont
  {Pouquet}}\ and\ \bibinfo {author} {\bibfnamefont {G.~S.}\ \bibnamefont
  {Patterson}},\ }\href@noop {} {\bibfield  {journal} {\bibinfo  {journal} {J.
  Fluid Mech.}\ }\textbf {\bibinfo {volume} {85}},\ \bibinfo {pages} {305}
  (\bibinfo {year} {1978})}\BibitemShut {NoStop}%
\bibitem [{\citenamefont {Biskamp}(1993)}]{Biskamp93}%
  \BibitemOpen
  \bibfield  {author} {\bibinfo {author} {\bibfnamefont {D.}~\bibnamefont
  {Biskamp}},\ }\href@noop {} {\emph {\bibinfo {title} {{Nonlinear
  Magnetohydrodynamics.}}}},\ \bibinfo {edition} {1st}\ ed.\ (\bibinfo
  {publisher} {Cambridge University Press},\ \bibinfo {year}
  {1993})\BibitemShut {NoStop}%
\bibitem [{\citenamefont {Alexakis}(2013)}]{Alexakis13}%
  \BibitemOpen
  \bibfield  {author} {\bibinfo {author} {\bibfnamefont {A.}~\bibnamefont
  {Alexakis}},\ }\href@noop {} {\bibfield  {journal} {\bibinfo  {journal}
  {Phys. Rev. Lett.}\ }\textbf {\bibinfo {volume} {110}},\ \bibinfo {pages}
  {084502} (\bibinfo {year} {2013})}\BibitemShut {NoStop}%
\bibitem [{\citenamefont {Sreenivasan}(1984)}]{Sreenivasan84}%
  \BibitemOpen
  \bibfield  {author} {\bibinfo {author} {\bibfnamefont {K.~R.}\ \bibnamefont
  {Sreenivasan}},\ }\href@noop {} {\bibfield  {journal} {\bibinfo  {journal}
  {Phys. Fluids}\ }\textbf {\bibinfo {volume} {27}},\ \bibinfo {pages} {1048}
  (\bibinfo {year} {1984})}\BibitemShut {NoStop}%
\bibitem [{\citenamefont {Sreenivasan}(1998)}]{Sreenivasan98}%
  \BibitemOpen
  \bibfield  {author} {\bibinfo {author} {\bibfnamefont {K.~R.}\ \bibnamefont
  {Sreenivasan}},\ }\href@noop {} {\bibfield  {journal} {\bibinfo  {journal}
  {Phys. Fluids}\ }\textbf {\bibinfo {volume} {10}},\ \bibinfo {pages} {528}
  (\bibinfo {year} {1998})}\BibitemShut {NoStop}%
\bibitem [{\citenamefont {McComb}(2014)}]{McComb14a}%
  \BibitemOpen
  \bibfield  {author} {\bibinfo {author} {\bibfnamefont {W.~D.}\ \bibnamefont
  {McComb}},\ }\href@noop {} {\emph {\bibinfo {title} {{Homogeneous,
  {I}sotropic {T}urbulence: {P}henomenology, {R}enormalization and
  {S}tatistical {C}losures}}}}\ (\bibinfo  {publisher} {Oxford University
  Press},\ \bibinfo {year} {2014})\BibitemShut {NoStop}%
\bibitem [{\citenamefont {McComb}\ \emph
  {et~al.}(2015{\natexlab{a}})\citenamefont {McComb}, \citenamefont {Berera},
  \citenamefont {Yoffe},\ and\ \citenamefont {Linkmann}}]{McComb15a}%
  \BibitemOpen
  \bibfield  {author} {\bibinfo {author} {\bibfnamefont {W.~D.}\ \bibnamefont
  {McComb}}, \bibinfo {author} {\bibfnamefont {A.}~\bibnamefont {Berera}},
  \bibinfo {author} {\bibfnamefont {S.~R.}\ \bibnamefont {Yoffe}}, \ and\
  \bibinfo {author} {\bibfnamefont {M.~F.}\ \bibnamefont {Linkmann}},\
  }\href@noop {} {\bibfield  {journal} {\bibinfo  {journal} {Phys. Rev. E}\
  }\textbf {\bibinfo {volume} {91}},\ \bibinfo {pages} {{043013}} (\bibinfo
  {year} {2015}{\natexlab{a}})}\BibitemShut {NoStop}%
\bibitem [{\citenamefont {Yeung}\ \emph {et~al.}(2015)\citenamefont {Yeung},
  \citenamefont {Zhai},\ and\ \citenamefont {Sreenivasan}}]{Yeung15}%
  \BibitemOpen
  \bibfield  {author} {\bibinfo {author} {\bibfnamefont {P.~K.}\ \bibnamefont
  {Yeung}}, \bibinfo {author} {\bibfnamefont {X.~M.}\ \bibnamefont {Zhai}}, \
  and\ \bibinfo {author} {\bibfnamefont {K.~R.}\ \bibnamefont {Sreenivasan}},\
  }\href@noop {} {\bibfield  {journal} {\bibinfo  {journal} {PNAS}\ }\textbf
  {\bibinfo {volume} {112}},\ \bibinfo {pages} {12633–12638} (\bibinfo {year}
  {2015})}\BibitemShut {NoStop}%
\bibitem [{\citenamefont {Jagannathan}\ and\ \citenamefont
  {Donzis}(2016)}]{Donzis16}%
  \BibitemOpen
  \bibfield  {author} {\bibinfo {author} {\bibfnamefont {S.}~\bibnamefont
  {Jagannathan}}\ and\ \bibinfo {author} {\bibfnamefont {D.~A.}\ \bibnamefont
  {Donzis}},\ }\href@noop {} {\bibfield  {journal} {\bibinfo  {journal} {J.
  Fluid Mech.}\ }\textbf {\bibinfo {volume} {789}},\ \bibinfo {pages} {669}
  (\bibinfo {year} {2016})}\BibitemShut {NoStop}%
\bibitem [{\citenamefont {Mininni}\ and\ \citenamefont
  {Pouquet}(2009)}]{Mininni09}%
  \BibitemOpen
  \bibfield  {author} {\bibinfo {author} {\bibfnamefont {P.~D.}\ \bibnamefont
  {Mininni}}\ and\ \bibinfo {author} {\bibfnamefont {A.~G.}\ \bibnamefont
  {Pouquet}},\ }\href@noop {} {\bibfield  {journal} {\bibinfo  {journal} {Phys.
  Rev. E.}\ }\textbf {\bibinfo {volume} {80}},\ \bibinfo {pages} {025401}
  (\bibinfo {year} {2009})}\BibitemShut {NoStop}%
\bibitem [{\citenamefont {Dallas}\ and\ \citenamefont
  {Alexakis}(2014)}]{Dallas14b}%
  \BibitemOpen
  \bibfield  {author} {\bibinfo {author} {\bibfnamefont {V.}~\bibnamefont
  {Dallas}}\ and\ \bibinfo {author} {\bibfnamefont {A.}~\bibnamefont
  {Alexakis}},\ }\href@noop {} {\bibfield  {journal} {\bibinfo  {journal}
  {Astrophys. J.}\ }\textbf {\bibinfo {volume} {788}},\ \bibinfo {pages} {L36}
  (\bibinfo {year} {2014})}\BibitemShut {NoStop}%
\bibitem [{\citenamefont {Linkmann}\ \emph {et~al.}(2015)\citenamefont
  {Linkmann}, \citenamefont {Berera}, \citenamefont {McComb},\ and\
  \citenamefont {McKay}}]{Linkmann15a}%
  \BibitemOpen
  \bibfield  {author} {\bibinfo {author} {\bibfnamefont {M.~F.}\ \bibnamefont
  {Linkmann}}, \bibinfo {author} {\bibfnamefont {A.}~\bibnamefont {Berera}},
  \bibinfo {author} {\bibfnamefont {W.~D.}\ \bibnamefont {McComb}}, \ and\
  \bibinfo {author} {\bibfnamefont {M.~E.}\ \bibnamefont {McKay}},\ }\href@noop
  {} {\bibfield  {journal} {\bibinfo  {journal} {Phys. Rev. Lett.}\ }\textbf
  {\bibinfo {volume} {114}},\ \bibinfo {pages} {235001} (\bibinfo {year}
  {2015})}\BibitemShut {NoStop}%
\bibitem [{\citenamefont {Politano}\ and\ \citenamefont
  {Pouquet}(1998)}]{Politano98}%
  \BibitemOpen
  \bibfield  {author} {\bibinfo {author} {\bibfnamefont {H.}~\bibnamefont
  {Politano}}\ and\ \bibinfo {author} {\bibfnamefont {A.}~\bibnamefont
  {Pouquet}},\ }\href@noop {} {\bibfield  {journal} {\bibinfo  {journal} {Phys.
  Rev. E}\ }\textbf {\bibinfo {volume} {57}},\ \bibinfo {pages} {R21} (\bibinfo
  {year} {1998})}\BibitemShut {NoStop}%
\bibitem [{\citenamefont {Els{\"a}sser}(1950)}]{Elsasser50}%
  \BibitemOpen
  \bibfield  {author} {\bibinfo {author} {\bibfnamefont {W.~M.}\ \bibnamefont
  {Els{\"a}sser}},\ }\href@noop {} {\bibfield  {journal} {\bibinfo  {journal}
  {Phys. Rev.}\ }\textbf {\bibinfo {volume} {79}},\ \bibinfo {pages} {183}
  (\bibinfo {year} {1950})}\BibitemShut {NoStop}%
\bibitem [{\citenamefont {Chandrasekhar}(1951)}]{Chandrasekhar51}%
  \BibitemOpen
  \bibfield  {author} {\bibinfo {author} {\bibfnamefont {S.}~\bibnamefont
  {Chandrasekhar}},\ }\href@noop {} {\bibfield  {journal} {\bibinfo  {journal}
  {Proc. Roy. Soc. London. Series A}\ }\textbf {\bibinfo {volume} {204}},\
  \bibinfo {pages} {435} (\bibinfo {year} {1951})}\BibitemShut {NoStop}%
\bibitem [{Note1()}]{Note1}%
  \BibitemOpen
  \bibinfo {note} {The scaling is ill-defined for the (measure zero) cases
  $\protect \bm {u} = \pm \protect \bm {b}$, which correspond to exact
  solutions to the MHD equations where the nonlinear terms vanish. Thus no
  turbulent transfer is possible, and these cases are not amenable to an
  analysis which assumes nonzero energy transfer \cite
  {Politano98}.}\BibitemShut {Stop}%
\bibitem [{\citenamefont {Novikov}(1965)}]{Novikov65}%
  \BibitemOpen
  \bibfield  {author} {\bibinfo {author} {\bibfnamefont {E.~A.}\ \bibnamefont
  {Novikov}},\ }\href@noop {} {\bibfield  {journal} {\bibinfo  {journal}
  {Soviet Physics JETP}\ }\textbf {\bibinfo {volume} {20}},\ \bibinfo {pages}
  {1290} (\bibinfo {year} {1965})}\BibitemShut {NoStop}%
\bibitem [{\citenamefont {Lundgren}(2002)}]{Lundgren02}%
  \BibitemOpen
  \bibfield  {author} {\bibinfo {author} {\bibfnamefont {T.~S.}\ \bibnamefont
  {Lundgren}},\ }\href@noop {} {\bibfield  {journal} {\bibinfo  {journal}
  {Phys. Fluids}\ }\textbf {\bibinfo {volume} {14}},\ \bibinfo {pages} {638}
  (\bibinfo {year} {2002})}\BibitemShut {NoStop}%
\bibitem [{\citenamefont {Linkmann}\ \emph {et~al.}(2016)\citenamefont
  {Linkmann}, \citenamefont {Berera}, \citenamefont {McKay},\ and\
  \citenamefont {J{\"a}ger}}]{Linkmann16a}%
  \BibitemOpen
  \bibfield  {author} {\bibinfo {author} {\bibfnamefont {M.~F.}\ \bibnamefont
  {Linkmann}}, \bibinfo {author} {\bibfnamefont {A.}~\bibnamefont {Berera}},
  \bibinfo {author} {\bibfnamefont {M.~E.}\ \bibnamefont {McKay}}, \ and\
  \bibinfo {author} {\bibfnamefont {J.}~\bibnamefont {J{\"a}ger}},\ }\href@noop
  {} {\bibfield  {journal} {\bibinfo  {journal} {J. Fluid Mech.}\ }\textbf
  {\bibinfo {volume} {791}},\ \bibinfo {pages} {61} (\bibinfo {year}
  {2016})}\BibitemShut {NoStop}%
\bibitem [{\citenamefont {Berera}\ and\ \citenamefont
  {Linkmann}(2014)}]{Berera14a}%
  \BibitemOpen
  \bibfield  {author} {\bibinfo {author} {\bibfnamefont {A.}~\bibnamefont
  {Berera}}\ and\ \bibinfo {author} {\bibfnamefont {M.~F.}\ \bibnamefont
  {Linkmann}},\ }\href@noop {} {\bibfield  {journal} {\bibinfo  {journal}
  {Phys. Rev. E}\ }\textbf {\bibinfo {volume} {90}},\ \bibinfo {pages}
  {041003(R)} (\bibinfo {year} {2014})}\BibitemShut {NoStop}%
\bibitem [{\citenamefont {Linkmann}(2016)}]{Linkmann_thesis}%
  \BibitemOpen
  \bibfield  {author} {\bibinfo {author} {\bibfnamefont {M.}~\bibnamefont
  {Linkmann}},\ }\emph {\bibinfo {title} {{Self-organisation in
  (magneto)hydrodynamic turbulence}}},\ \href@noop {} {Ph.D. thesis},\ \bibinfo
   {school} {{University of Edinburgh}} (\bibinfo {year} {{2016}})\BibitemShut
  {NoStop}%
\bibitem [{\citenamefont {Brandenburg}(2001)}]{Brandenburg01}%
  \BibitemOpen
  \bibfield  {author} {\bibinfo {author} {\bibfnamefont {A.}~\bibnamefont
  {Brandenburg}},\ }\href@noop {} {\bibfield  {journal} {\bibinfo  {journal}
  {Astrophys. J.}\ }\textbf {\bibinfo {volume} {550}},\ \bibinfo {pages} {824}
  (\bibinfo {year} {2001})}\BibitemShut {NoStop}%
\bibitem [{\citenamefont {Dallas}\ and\ \citenamefont
  {Alexakis}(2015)}]{Dallas14a}%
  \BibitemOpen
  \bibfield  {author} {\bibinfo {author} {\bibfnamefont {V.}~\bibnamefont
  {Dallas}}\ and\ \bibinfo {author} {\bibfnamefont {A.}~\bibnamefont
  {Alexakis}},\ }\href@noop {} {\bibfield  {journal} {\bibinfo  {journal}
  {Phys. Fluids}\ }\textbf {\bibinfo {volume} {27}},\ \bibinfo {pages} {045105}
  (\bibinfo {year} {2015})}\BibitemShut {NoStop}%
\bibitem [{\citenamefont {Sahoo}\ \emph {et~al.}(2011)\citenamefont {Sahoo},
  \citenamefont {Perlekar},\ and\ \citenamefont {Pandit}}]{Sahoo11}%
  \BibitemOpen
  \bibfield  {author} {\bibinfo {author} {\bibfnamefont {G.}~\bibnamefont
  {Sahoo}}, \bibinfo {author} {\bibfnamefont {P.}~\bibnamefont {Perlekar}}, \
  and\ \bibinfo {author} {\bibfnamefont {R.}~\bibnamefont {Pandit}},\
  }\href@noop {} {\bibfield  {journal} {\bibinfo  {journal} {New Journal of
  Physics}\ }\textbf {\bibinfo {volume} {13}},\ \bibinfo {pages} {013036}
  (\bibinfo {year} {2011})}\BibitemShut {NoStop}%
\bibitem [{\citenamefont {McComb}\ \emph
  {et~al.}(2015{\natexlab{b}})\citenamefont {McComb}, \citenamefont {Linkmann},
  \citenamefont {Berera}, \citenamefont {Yoffe},\ and\ \citenamefont
  {Jankauskas}}]{McComb15b}%
  \BibitemOpen
  \bibfield  {author} {\bibinfo {author} {\bibfnamefont {W.~D.}\ \bibnamefont
  {McComb}}, \bibinfo {author} {\bibfnamefont {M.~F.}\ \bibnamefont
  {Linkmann}}, \bibinfo {author} {\bibfnamefont {A.}~\bibnamefont {Berera}},
  \bibinfo {author} {\bibfnamefont {S.~R.}\ \bibnamefont {Yoffe}}, \ and\
  \bibinfo {author} {\bibfnamefont {B.}~\bibnamefont {Jankauskas}},\
  }\href@noop {} {\bibfield  {journal} {\bibinfo  {journal} {J. Phys. A: Math.
  Theor.}\ }\textbf {\bibinfo {volume} {48}},\ \bibinfo {pages} {25FT01}
  (\bibinfo {year} {2015}{\natexlab{b}})}\BibitemShut {NoStop}%
\bibitem [{\citenamefont {Linkmann}\ and\ \citenamefont
  {Morozov}(2015)}]{Linkmann15b}%
  \BibitemOpen
  \bibfield  {author} {\bibinfo {author} {\bibfnamefont {M.~F.}\ \bibnamefont
  {Linkmann}}\ and\ \bibinfo {author} {\bibfnamefont {A.}~\bibnamefont
  {Morozov}},\ }\href@noop {} {\bibfield  {journal} {\bibinfo  {journal} {Phys.
  Rev. Lett.}\ }\textbf {\bibinfo {volume} {115}},\ \bibinfo {pages} {134502}
  (\bibinfo {year} {2015})}\BibitemShut {NoStop}%
\bibitem [{\citenamefont {M\"uller}\ \emph {et~al.}(2012)\citenamefont
  {M\"uller}, \citenamefont {Malapaka},\ and\ \citenamefont
  {Busse}}]{Mueller12}%
  \BibitemOpen
  \bibfield  {author} {\bibinfo {author} {\bibfnamefont {W.~C.}\ \bibnamefont
  {M\"uller}}, \bibinfo {author} {\bibfnamefont {S.~K.}\ \bibnamefont
  {Malapaka}}, \ and\ \bibinfo {author} {\bibfnamefont {A.}~\bibnamefont
  {Busse}},\ }\href@noop {} {\bibfield  {journal} {\bibinfo  {journal} {Phys.
  Rev. E}\ }\textbf {\bibinfo {volume} {85}},\ \bibinfo {pages} {015302}
  (\bibinfo {year} {2012})}\BibitemShut {NoStop}%
\bibitem [{\citenamefont {Malapaka}\ and\ \citenamefont
  {M\"uller}(2013)}]{Malapaka13}%
  \BibitemOpen
  \bibfield  {author} {\bibinfo {author} {\bibfnamefont {S.~K.}\ \bibnamefont
  {Malapaka}}\ and\ \bibinfo {author} {\bibfnamefont {W.-C.}\ \bibnamefont
  {M\"uller}},\ }\href@noop {} {\bibfield  {journal} {\bibinfo  {journal}
  {Astrophys. J.}\ }\textbf {\bibinfo {volume} {778}},\ \bibinfo {pages} {21}
  (\bibinfo {year} {2013})}\BibitemShut {NoStop}%
\bibitem [{\citenamefont {Linkmann}\ and\ \citenamefont
  {Dallas}(2016)}]{Linkmann16b}%
  \BibitemOpen
  \bibfield  {author} {\bibinfo {author} {\bibfnamefont {M.}~\bibnamefont
  {Linkmann}}\ and\ \bibinfo {author} {\bibfnamefont {V.}~\bibnamefont
  {Dallas}},\ }\href@noop {} {\bibfield  {journal} {\bibinfo  {journal} {Phys.
  Rev. E}\ }\textbf {\bibinfo {volume} {94}},\ \bibinfo {pages} {053209}
  (\bibinfo {year} {2016})}\BibitemShut {NoStop}%
\bibitem [{\citenamefont {{The data is publicly available}}()}]{data15a}%
  \BibitemOpen
  \bibfield  {author} {\bibinfo {author} {\bibnamefont {{The data is publicly
  available}}},\ }\href@noop {} {}\bibinfo {howpublished}
  {\url{http://dx.doi.org/10.7488/ds/247}}\BibitemShut {NoStop}%
\bibitem [{\citenamefont {Biferale}\ \emph {et~al.}(2004)\citenamefont
  {Biferale}, \citenamefont {Lanotte},\ and\ \citenamefont
  {Toschi}}]{Biferale04}%
  \BibitemOpen
  \bibfield  {author} {\bibinfo {author} {\bibfnamefont {L.}~\bibnamefont
  {Biferale}}, \bibinfo {author} {\bibfnamefont {A.~S.}\ \bibnamefont
  {Lanotte}}, \ and\ \bibinfo {author} {\bibfnamefont {F.}~\bibnamefont
  {Toschi}},\ }\href@noop {} {\bibfield  {journal} {\bibinfo  {journal} {Phys.
  Rev. Lett.}\ }\textbf {\bibinfo {volume} {92}},\ \bibinfo {pages} {094503}
  (\bibinfo {year} {2004})}\BibitemShut {NoStop}%
\bibitem [{\citenamefont {Mitra}\ \emph {et~al.}(2005)\citenamefont {Mitra},
  \citenamefont {Bec}, \citenamefont {Pandit},\ and\ \citenamefont
  {Frisch}}]{Mitra05}%
  \BibitemOpen
  \bibfield  {author} {\bibinfo {author} {\bibfnamefont {D.}~\bibnamefont
  {Mitra}}, \bibinfo {author} {\bibfnamefont {J.}~\bibnamefont {Bec}}, \bibinfo
  {author} {\bibfnamefont {R.}~\bibnamefont {Pandit}}, \ and\ \bibinfo {author}
  {\bibfnamefont {U.}~\bibnamefont {Frisch}},\ }\href@noop {} {\bibfield
  {journal} {\bibinfo  {journal} {Phys. Rev. Lett.}\ }\textbf {\bibinfo
  {volume} {94}},\ \bibinfo {pages} {{194501}} (\bibinfo {year}
  {2005})}\BibitemShut {NoStop}%
\bibitem [{\citenamefont {Brandenburg}(2014)}]{Brandenburg14b}%
  \BibitemOpen
  \bibfield  {author} {\bibinfo {author} {\bibfnamefont {A.}~\bibnamefont
  {Brandenburg}},\ }\href@noop {} {\bibfield  {journal} {\bibinfo  {journal}
  {ApJ}\ }\textbf {\bibinfo {volume} {791}},\ \bibinfo {pages} {12} (\bibinfo
  {year} {2014})}\BibitemShut {NoStop}%
\bibitem [{\citenamefont {Jim{\'e}nez}\ \emph {et~al.}(1993)\citenamefont
  {Jim{\'e}nez}, \citenamefont {Wray}, \citenamefont {Saffman},\ and\
  \citenamefont {Rogallo}}]{Jimenez93}%
  \BibitemOpen
  \bibfield  {author} {\bibinfo {author} {\bibfnamefont {J.}~\bibnamefont
  {Jim{\'e}nez}}, \bibinfo {author} {\bibfnamefont {A.~A.}\ \bibnamefont
  {Wray}}, \bibinfo {author} {\bibfnamefont {P.~G.}\ \bibnamefont {Saffman}}, \
  and\ \bibinfo {author} {\bibfnamefont {R.~S.}\ \bibnamefont {Rogallo}},\
  }\href@noop {} {\bibfield  {journal} {\bibinfo  {journal} {J. Fluid Mech.}\
  }\textbf {\bibinfo {volume} {255}},\ \bibinfo {pages} {65} (\bibinfo {year}
  {1993})}\BibitemShut {NoStop}%
\bibitem [{\citenamefont {Wang}\ \emph {et~al.}(1996)\citenamefont {Wang},
  \citenamefont {Chen}, \citenamefont {Brasseur},\ and\ \citenamefont
  {Wyngaard}}]{Wang96}%
  \BibitemOpen
  \bibfield  {author} {\bibinfo {author} {\bibfnamefont {L.-P.}\ \bibnamefont
  {Wang}}, \bibinfo {author} {\bibfnamefont {S.}~\bibnamefont {Chen}}, \bibinfo
  {author} {\bibfnamefont {J.~G.}\ \bibnamefont {Brasseur}}, \ and\ \bibinfo
  {author} {\bibfnamefont {J.~C.}\ \bibnamefont {Wyngaard}},\ }\href@noop {}
  {\bibfield  {journal} {\bibinfo  {journal} {J. Fluid Mech.}\ }\textbf
  {\bibinfo {volume} {309}},\ \bibinfo {pages} {113} (\bibinfo {year}
  {1996})}\BibitemShut {NoStop}%
\bibitem [{\citenamefont {Yeung}\ and\ \citenamefont {Zhou}(1997)}]{Yeung97}%
  \BibitemOpen
  \bibfield  {author} {\bibinfo {author} {\bibfnamefont {P.~K.}\ \bibnamefont
  {Yeung}}\ and\ \bibinfo {author} {\bibfnamefont {Y.}~\bibnamefont {Zhou}},\
  }\href@noop {} {\bibfield  {journal} {\bibinfo  {journal} {Phys. Rev. E}\
  }\textbf {\bibinfo {volume} {56}},\ \bibinfo {pages} {1746} (\bibinfo {year}
  {1997})}\BibitemShut {NoStop}%
\bibitem [{\citenamefont {Cao}\ \emph {et~al.}(1999)\citenamefont {Cao},
  \citenamefont {Chen},\ and\ \citenamefont {Doolen}}]{Cao99}%
  \BibitemOpen
  \bibfield  {author} {\bibinfo {author} {\bibfnamefont {N.}~\bibnamefont
  {Cao}}, \bibinfo {author} {\bibfnamefont {S.}~\bibnamefont {Chen}}, \ and\
  \bibinfo {author} {\bibfnamefont {G.~D.}\ \bibnamefont {Doolen}},\
  }\href@noop {} {\bibfield  {journal} {\bibinfo  {journal} {Phys. Fluids}\
  }\textbf {\bibinfo {volume} {11}},\ \bibinfo {pages} {2235} (\bibinfo {year}
  {1999})}\BibitemShut {NoStop}%
\bibitem [{\citenamefont {Pearson}\ \emph {et~al.}(2002)\citenamefont
  {Pearson}, \citenamefont {Krogstad},\ and\ \citenamefont
  {{van~de~Water}}}]{Pearson02}%
  \BibitemOpen
  \bibfield  {author} {\bibinfo {author} {\bibfnamefont {B.~R.}\ \bibnamefont
  {Pearson}}, \bibinfo {author} {\bibfnamefont {P.~A.}\ \bibnamefont
  {Krogstad}}, \ and\ \bibinfo {author} {\bibfnamefont {W.}~\bibnamefont
  {{van~de~Water}}},\ }\href@noop {} {\bibfield  {journal} {\bibinfo  {journal}
  {Phys. Fluids}\ }\textbf {\bibinfo {volume} {14}},\ \bibinfo {pages} {1288}
  (\bibinfo {year} {2002})}\BibitemShut {NoStop}%
\bibitem [{\citenamefont {Kaneda}\ \emph {et~al.}(2003)\citenamefont {Kaneda},
  \citenamefont {Ishihara}, \citenamefont {Yokokawa}, \citenamefont {Itakura},\
  and\ \citenamefont {Uno}}]{Kaneda03}%
  \BibitemOpen
  \bibfield  {author} {\bibinfo {author} {\bibfnamefont {Y.}~\bibnamefont
  {Kaneda}}, \bibinfo {author} {\bibfnamefont {T.}~\bibnamefont {Ishihara}},
  \bibinfo {author} {\bibfnamefont {M.}~\bibnamefont {Yokokawa}}, \bibinfo
  {author} {\bibfnamefont {K.}~\bibnamefont {Itakura}}, \ and\ \bibinfo
  {author} {\bibfnamefont {A.}~\bibnamefont {Uno}},\ }\href@noop {} {\bibfield
  {journal} {\bibinfo  {journal} {Phys. Fluids}\ }\textbf {\bibinfo {volume}
  {15}},\ \bibinfo {pages} {L21} (\bibinfo {year} {2003})}\BibitemShut
  {NoStop}%
\bibitem [{\citenamefont {Pearson}\ \emph {et~al.}(2004)\citenamefont
  {Pearson}, \citenamefont {Yousef}, \citenamefont {Haugen}, \citenamefont
  {Brandenburg},\ and\ \citenamefont {Krogstad}}]{Pearson04a}%
  \BibitemOpen
  \bibfield  {author} {\bibinfo {author} {\bibfnamefont {B.~R.}\ \bibnamefont
  {Pearson}}, \bibinfo {author} {\bibfnamefont {T.~A.}\ \bibnamefont {Yousef}},
  \bibinfo {author} {\bibfnamefont {N.~E.~L.}\ \bibnamefont {Haugen}}, \bibinfo
  {author} {\bibfnamefont {A.}~\bibnamefont {Brandenburg}}, \ and\ \bibinfo
  {author} {\bibfnamefont {P.~A.}\ \bibnamefont {Krogstad}},\ }\href@noop {}
  {\bibfield  {journal} {\bibinfo  {journal} {Phys. Rev. E}\ }\textbf {\bibinfo
  {volume} {70}},\ \bibinfo {pages} {056301} (\bibinfo {year}
  {2004})}\BibitemShut {NoStop}%
\bibitem [{\citenamefont {Donzis}\ \emph {et~al.}(2005)\citenamefont {Donzis},
  \citenamefont {Sreenivasan},\ and\ \citenamefont {Yeung}}]{Donzis05}%
  \BibitemOpen
  \bibfield  {author} {\bibinfo {author} {\bibfnamefont {D.~A.}\ \bibnamefont
  {Donzis}}, \bibinfo {author} {\bibfnamefont {K.~R.}\ \bibnamefont
  {Sreenivasan}}, \ and\ \bibinfo {author} {\bibfnamefont {P.~K.}\ \bibnamefont
  {Yeung}},\ }\href@noop {} {\bibfield  {journal} {\bibinfo  {journal} {J.
  Fluid Mech.}\ }\textbf {\bibinfo {volume} {532}},\ \bibinfo {pages} {199}
  (\bibinfo {year} {2005})}\BibitemShut {NoStop}%
\bibitem [{\citenamefont {Bos}\ \emph {et~al.}(2007)\citenamefont {Bos},
  \citenamefont {Shao},\ and\ \citenamefont {Bertoglio}}]{Bos07}%
  \BibitemOpen
  \bibfield  {author} {\bibinfo {author} {\bibfnamefont {W.~J.~T.}\
  \bibnamefont {Bos}}, \bibinfo {author} {\bibfnamefont {L.}~\bibnamefont
  {Shao}}, \ and\ \bibinfo {author} {\bibfnamefont {J.-P.}\ \bibnamefont
  {Bertoglio}},\ }\href@noop {} {\bibfield  {journal} {\bibinfo  {journal}
  {Phys. Fluids}\ }\textbf {\bibinfo {volume} {19}},\ \bibinfo {pages} {045101}
  (\bibinfo {year} {2007})}\BibitemShut {NoStop}%
\bibitem [{\citenamefont {Yeung}\ \emph {et~al.}(2012)\citenamefont {Yeung},
  \citenamefont {Donzis},\ and\ \citenamefont {Sreenivasan}}]{Yeung12}%
  \BibitemOpen
  \bibfield  {author} {\bibinfo {author} {\bibfnamefont {P.~K.}\ \bibnamefont
  {Yeung}}, \bibinfo {author} {\bibfnamefont {D.~A.}\ \bibnamefont {Donzis}}, \
  and\ \bibinfo {author} {\bibfnamefont {K.~R.}\ \bibnamefont {Sreenivasan}},\
  }\href@noop {} {\bibfield  {journal} {\bibinfo  {journal} {J. Fluid Mech.}\
  }\textbf {\bibinfo {volume} {700}},\ \bibinfo {pages} {5} (\bibinfo {year}
  {2012})}\BibitemShut {NoStop}%
\bibitem [{\citenamefont {Doering}(2009)}]{Doering09}%
  \BibitemOpen
  \bibfield  {author} {\bibinfo {author} {\bibfnamefont {C.~R.}\ \bibnamefont
  {Doering}},\ }\href@noop {} {\bibfield  {journal} {\bibinfo  {journal} {Annu.
  Rev. Fl. Mech.}\ }\textbf {\bibinfo {volume} {41}},\ \bibinfo {pages} {109}
  (\bibinfo {year} {2009})}\BibitemShut {NoStop}%
\bibitem [{\citenamefont {Mazzi}\ and\ \citenamefont
  {Vassilicos}(2004)}]{Mazzi04}%
  \BibitemOpen
  \bibfield  {author} {\bibinfo {author} {\bibfnamefont {B.}~\bibnamefont
  {Mazzi}}\ and\ \bibinfo {author} {\bibfnamefont {J.~C.}\ \bibnamefont
  {Vassilicos}},\ }\href@noop {} {\bibfield  {journal} {\bibinfo  {journal} {J.
  Fluid Mech.}\ }\textbf {\bibinfo {volume} {502}},\ \bibinfo {pages} {65}
  (\bibinfo {year} {2004})}\BibitemShut {NoStop}%
\bibitem [{\citenamefont {Valente}\ \emph {et~al.}(2014)\citenamefont
  {Valente}, \citenamefont {Onishi},\ and\ \citenamefont {{da
  Silva}}}]{Valente14}%
  \BibitemOpen
  \bibfield  {author} {\bibinfo {author} {\bibfnamefont {P.~C.}\ \bibnamefont
  {Valente}}, \bibinfo {author} {\bibfnamefont {R.}~\bibnamefont {Onishi}}, \
  and\ \bibinfo {author} {\bibfnamefont {C.~B.}\ \bibnamefont {{da Silva}}},\
  }\href@noop {} {\bibfield  {journal} {\bibinfo  {journal} {Phys. Rev. E}\
  }\textbf {\bibinfo {volume} {90}},\ \bibinfo {pages} {023003} (\bibinfo
  {year} {2014})}\BibitemShut {NoStop}%
\bibitem [{\citenamefont {Vassilicos}(2015)}]{Vassilicos15}%
  \BibitemOpen
  \bibfield  {author} {\bibinfo {author} {\bibfnamefont {J.~C.}\ \bibnamefont
  {Vassilicos}},\ }\href@noop {} {\bibfield  {journal} {\bibinfo  {journal}
  {Annu. Rev. Fluid Mech.}\ }\textbf {\bibinfo {volume} {47}},\ \bibinfo
  {pages} {95} (\bibinfo {year} {2015})}\BibitemShut {NoStop}%
\bibitem [{\citenamefont {Matthaeus}\ and\ \citenamefont
  {Goldstein}(1982)}]{Matthaeus82a}%
  \BibitemOpen
  \bibfield  {author} {\bibinfo {author} {\bibfnamefont {W.~H.}\ \bibnamefont
  {Matthaeus}}\ and\ \bibinfo {author} {\bibfnamefont {M.~L.}\ \bibnamefont
  {Goldstein}},\ }\href@noop {} {\bibfield  {journal} {\bibinfo  {journal} {J.
  Geophys. Res.}\ }\textbf {\bibinfo {volume} {87}},\ \bibinfo {pages} {6011}
  (\bibinfo {year} {1982})}\BibitemShut {NoStop}%
\bibitem [{\citenamefont {Wan}\ \emph {et~al.}(2009)\citenamefont {Wan},
  \citenamefont {Servidio}, \citenamefont {Oughton},\ and\ \citenamefont
  {Matthaeus}}]{Wan09}%
  \BibitemOpen
  \bibfield  {author} {\bibinfo {author} {\bibfnamefont {M.}~\bibnamefont
  {Wan}}, \bibinfo {author} {\bibfnamefont {S.}~\bibnamefont {Servidio}},
  \bibinfo {author} {\bibfnamefont {S.}~\bibnamefont {Oughton}}, \ and\
  \bibinfo {author} {\bibfnamefont {W.~H.}\ \bibnamefont {Matthaeus}},\
  }\href@noop {} {\bibfield  {journal} {\bibinfo  {journal} {Phys. Plasmas}\
  }\textbf {\bibinfo {volume} {16}},\ \bibinfo {pages} {090703} (\bibinfo
  {year} {2009})}\BibitemShut {NoStop}%
\bibitem [{\citenamefont {Berger}(1997)}]{Berger97}%
  \BibitemOpen
  \bibfield  {author} {\bibinfo {author} {\bibfnamefont {M.~A.}\ \bibnamefont
  {Berger}},\ }\href@noop {} {\bibfield  {journal} {\bibinfo  {journal} {J.
  Geophys. Res.}\ }\textbf {\bibinfo {volume} {102}},\ \bibinfo {pages} {2637}
  (\bibinfo {year} {1997})}\BibitemShut {NoStop}%
\bibitem [{\citenamefont {Berger}(1999)}]{Berger99}%
  \BibitemOpen
  \bibfield  {author} {\bibinfo {author} {\bibfnamefont {M.~A.}\ \bibnamefont
  {Berger}},\ }\href@noop {} {\bibfield  {journal} {\bibinfo  {journal} {Plasma
  Physics and Controlled Fusion}\ }\textbf {\bibinfo {volume} {41}},\ \bibinfo
  {pages} {B167} (\bibinfo {year} {1999})}\BibitemShut {NoStop}%
\end{thebibliography}%

\end{document}